\documentclass{article}
\usepackage{arxiv}

\pdfoutput=1

\usepackage[utf8]{inputenc} 
\usepackage[T1]{fontenc}    
\usepackage{url}   
\usepackage{booktabs} 
\usepackage{amsfonts} 
\usepackage{nicefrac} 
\usepackage{microtype}
\usepackage{lipsum}
\usepackage{graphicx}
\graphicspath{ {./figures/} }
\usepackage[version=4]{mhchem} 
\usepackage{tabularx}
\usepackage{adjustbox}
\usepackage{subcaption}

\usepackage{siunitx}
\usepackage{multirow}
\usepackage[table]{xcolor}
\usepackage{listings} 
\usepackage{xcolor} 
\usepackage{amsmath}
\usepackage{amssymb}
\usepackage{amsthm}
\usepackage{mathtools}
\usepackage{tikz}
\usetikzlibrary{matrix,chains,positioning,decorations.pathreplacing,arrows}
\usepackage{algorithm}
\usepackage[normalem]{ulem}
\usepackage[noend]{algpseudocode}
\usepackage{filecontents}
\usepackage{xr-hyper}
\usepackage{hyperref} 
\usepackage{cleveref}
\usepackage[numbers, sort&compress]{natbib}

\makeatletter
\newcommand*{\addFileDependency}[1]{
  \typeout{(#1)}
  \@addtofilelist{#1}
  \IfFileExists{#1}{}{\ypeout{No file #1.}}
}
\makeatother

\makeatletter
\newcommand*{\addAuxFileDependency}[1]{
  \makeatletter\@input{x#1.tex}\makeatother
}
\makeatother

\small

\numberwithin{equation}{section}
\numberwithin{equation}{subsection}

\algrenewcommand{\algorithmicrequire}{\textbf{Input:}}
\algrenewcommand{\algorithmicensure}{\textbf{Output:}}

\crefname{equation}{Eq.}{Eqs.}
\Crefname{equation}{Equation}{Equations}
\crefname{table}{Table}{Tables}
\Crefname{table}{Table}{Tables}
\crefname{figure}{Fig.}{Figs.}
\Crefname{figure}{Figure}{Figures}

\title{Modeling Batch Crystallization under Uncertainty Using Physics-informed Machine Learning}

\author{
  Dingqi Nai\textsuperscript{a}, Huayu Li\textsuperscript{b}, Martha Grover\textsuperscript{a}, Andrew J. Medford\textsuperscript{a}$^*$\\\\
  \textsuperscript{a}School of Chemical \& Biomolecular Engineering\\
  Georgia Institute of Technology\\
  Atlanta, GA 30332\\
  \textsuperscript{b}Material and Analytical Sciences\\
  Boehringer Ingelheim Pharmaceuticals, Inc.\\
  Ridgefield, CT 06877\\
}

\begin{document}

\maketitle

\let\thefootnote\relax\footnotetext{$*$~Corresponding author. Email: ajm@gatech.edu}

\begin{abstract}
The development of robust and reliable modeling approaches for crystallization processes is often challenging because of non-idealities in real data arising from various sources of uncertainty. This study investigated the effectiveness of physics-informed recurrent neural networks (PIRNNs) that integrate the mechanistic population balance model with recurrent neural networks under the presence of systematic and model uncertainties. Such uncertainties are represented by using synthetic data containing controlled noise, solubility shift, and limited sampling. The research demonstrates that PIRNNs achieve strong generalization and physical consistency, maintain stable learning behavior, and accurately recover kinetic parameters despite significant stochastic variations in the training data. In the case of systematic errors in the solubility model, the inclusion of physics regularization improved the test performance by more than an order of magnitude compared to purely data-driven models, whereas excessive weighting of physics increased error arising due to the model mismatch. The results also show that PIRNNs are able to recover model parameters and replicate crystallization dynamics even in the limit of very low sampling resolution. These findings validate the robustness of physics-informed machine learning in handling data imperfections and incomplete domain knowledge, providing a potential pathway toward reliable and practical hybrid modeling of crystallization dynamics and industrial process monitoring and control.


\end{abstract}

\section{Introduction} \label{sec:intro}

Crystallization is a fundamental process for the separation, purification, and production of solid materials in various industries, including pharmaceuticals, fine chemicals, and food processing \cite{Ahn2022SecondaryKinetics, Burcham2024PharmaceuticalProcess, Tavare1995IndustrialCrystallization}. In particular, in the pharmaceutical industries, more than 80\% of solid products involve at least one crystallization step, and over 90\% of the active pharmaceutical ingredients produced are crystals \cite{AbuBakar2009TheProcesses, Chen2011PharmaceuticalCrystallization, Lima2024NeuralCrystallization, Tang2021FormModeling}. The significance of crystallization on product characteristics, such as purity, particle size distribution, and polymorphic forms, makes understanding and precise control of this process necessary to consistently produce products that meet the quality specification \cite{Urwin2020ADevelopment}. Consequently, significant research has focused on developing deeper and more quantitative insights into crystallization process dynamics \cite{Kim2023ModelingStrategy, Lima2025ApplicationsPerspectives, Ulrich2013ProblemsCrystallization, Yu2007RecentControl}.

In recent decades, the advancement of process analytical technology (PAT) has revolutionized the understanding of crystallization process development by allowing real-time monitoring of this dynamic system \cite{Wu2015AnPerspective, Yu2004ApplicationsProcesses}. PAT tools, such as focused beam reflection measurement (FBRM), attenuated total reflection Fourier transform infrared (ATR-FTIR) spectroscopy, Raman spectroscopy, and in-situ imaging, provide detailed insights into crystallization dynamics including supersaturation and crystal size distribution (CSD) \cite{Kim2023ModelingStrategy, Li2013DevelopmentMeasurements, McDonald2021ReactiveReview, Roy2013PolymorphSpectroscopy, Salami2021InProduction}. 
However, the information gathered from PAT is often uncertain and imperfect, and effectively using these data for process control remains challenging without an appropriate process model that can interpret and relate the measurements to underlying system states \cite{Farouk2023ResolvingImprovement, Nagy2012AdvancesControl, Samad2013IntroducingProcesses}.

Physics-based crystallization modeling has traditionally formed the foundation for crystallization research and development. Based on the population balance model (PBM) and often coupled with mass and energy balances, physics-based models provide a rigorous mathematical description of nucleation, crystal growth, dissolution, aggregation, and breakage phenomena \cite{Kim2023ModelingStrategy, Omar2017CrystalReview, Rasmuson2019CrystallizationModeling, Szilagyi2021Cross-PharmaCrySiV}. The theoretical foundation and extrapolation ability grant physics-based models their own unique strength. However, parameter estimation of these models can be challenging due to the highly nonlinear structure of the equations and the requirement for a large amount of experimental data \cite{Bari2018SequentialKinetics, Besenhard2015EvaluationEquations}. Despite these challenges, physics-based models remain essential for process scale-up and design, as well as for gaining fundamental mechanistic understanding of the crystallization phenomenon.

Rapid developments in machine learning algorithms and the availability of extensive experimental datasets have also driven the appearance of purely data-driven modeling approaches \cite{Lima2025ApplicationsPerspectives, Lu2024RecentControl}. Techniques such as artificial neural networks (ANN), support vector machines, random forests, and other deep learning frameworks have been successfully applied to capture ``black-box'' empirical relationships between crystallization dynamics and input process variables \cite{An2022Data-drivenProcess, Daosud2017NeuralProcess, Kittisupakorn2017ImprovingCrystallizer,Li2014ModelProcess, Lima2023AnZones, Lima2024NeuralCrystallization, Ma2020ArtificialCarbonate, Manee2019ACharacterization, Nyande2024DatadrivenKinetics,VasanthKumar2008ModellingAnalysis, Wilkinson2022PredictingIntelligence}. Recent work has shown the efficiency of recurrent neural networks and time-series transformers for predicting crystallization dynamics and implementing neural network-based control strategies \cite{Lima2022DevelopmentProcess, Lima2024NeuralCrystallization, Sitapure2023ExploringShift, Sitapure2024MachineLSTMc, Zheng2022MachineProcess, Zheng2022OnlineDrift}. In contrast to physics-based models, data-driven models can handle large volumes of data without requiring explicit mechanistic knowledge. However, they often face inherent limitations of poor physical interpretability and extrapolation capabilities when dealing with novel systems or operating conditions beyond the training domain \cite{Griffin2016Data-DrivenCrystallization}.

Hybrid modeling, combining physics-based knowledge with data-driven techniques, has emerged as a promising framework to combine the strengths of both paradigms \cite{Bradley2022PerspectivesModeling, Schweidtmann2024AMethodologies}, with architectures like physics-informed neural networks (PINNs) being increasingly applied to various chemical engineering systems ranging from reaction kinetics and fluid dynamics to process control \cite{Alhajeri2022Physics-informedData, Asrav2023Physics-informedSystems, Atzori2025AEngineering, Batuwatta-Gamage2022ADrying, Cai2021Physics-informedReview, Choi2022Physics-informedDynamics, Gusmao2022Kinetics-informedNetworks, Ji2021Stiff-PINN:Kinetics, Koksal2024Physics-informedValidation, Moayedi2024Physics-InformedMismatch, Nai2024Micro-kineticNetworks, Shaw2024ModelingLearning, Wu2023TheEngineering, Xiao2023ModelingLearning, Zhu2024ABoilers}. This ``gray-box'' approach maintains the physical interpretability and extrapolation capabilities of physical models while using flexible machine learning models to handle large data volumes and discrepancies between simplified models and realistic systems, allowing better predictive performance under non-ideal conditions or sparse data regimes \cite{Galvanauskas2006DynamicModel, Georgieva2003Knowledge-basedPhenomena, Kittisupakorn2017ImprovingCrystallizer, Lima2023ImprovedEquations, Sitapure2023IntroducingCrystallization, Wu2023Physics-informedProcess}. Recent crystallization research has validated the advantages of hybrid modeling by showing the potential to achieve improved accuracy and robustness compared to purely data-driven approaches \cite{Lima2025ApplicationsPerspectives, Wu2023Physics-informedProcess}.

Nonetheless, practical challenges persist when applying these hybrid modeling approaches to real-world crystallization processes. Measurement noise, process disturbances, and other inherent system variability introduce two fundamental types of uncertainty: aleatoric uncertainty and epistemic uncertainty. Aleatoric uncertainty accounts for irreducible variations in observation coming from sources such as the stochastic nature of nucleation events, measurement device limitations, and random operational fluctuations. Epistemic uncertainty is more general, and arises from incomplete knowledge about the system and model, such as simplifying model assumptions, inconsistencies between the model and data, or limited coverage of experimental data \cite{Chan2024EstimatingModel, Hullermeier2021AleatoricMethods, Kiureghian2009AleatoryMatter, MatthiesQUANTIFYINGAPPLICATIONS, Segalman2013EpistemicModeling, Smith2025RethinkingUncertainty}. Previous studies have highlighted the effect of uncertainties on crystallization modeling using Monte Carlo procedures to propagate input uncertainties through PBMs, revealing significant impacts on predicted CSDs and process outcomes \cite{Samad2013IntroducingProcesses}. Recent work also attempts to quantify these uncertainties through Bayesian approaches, drawing more attention to understanding and accounting for these uncertainties when modeling crystallization processes \cite{Pessina2025IntegratedModels}.

This paper aims to extend these efforts by investigating the influence of data non-idealities on crystallization modeling using a hybrid approach building on the physics-informed recurrent neural networks (PIRNNs) recently developed by \citet{Wu2023Physics-informedProcess}. We focus on a relatively simple but representative case study of paracetamol batch cooling crystallization, where prior work has established a physical PBM model that we treat as the ``ground truth'' \cite{Li2017ModelingCrystallization} and  explore how aleatoric and several forms of epistemic uncertainty in synthetic ``observed data'' affect the performance and reliability of crystallization modeling. By systematically introducing controlled levels of measurement noise, model discrepancy, and sampling limitations we provide insight into the robustness and reliability of the proposed approach under imperfect data and incomplete domain knowledge. The results of this work show that hybrid PIRNNs models offer a robust option for parameterizing PBM models in the presence of aleatoric measurement noise and epistemic sampling uncertainty.
Moreover, we find that, when properly optimized, these hybrid models are able to flexibly account for epistemic uncertainty in solubility calibration curves and sampling frequency, outperforming both purely data-driven and physics-based models. These findings suggest that the PIRNNs approach is a promising strategy to create robust crystallization models that integrate practical PAT data with the known physicochemical governing mechanisms.

\section{Methodology} \label{sec:methods}
\subsection{Population Balance Model \& Method of Moments} \label{subsec:pbm}

In this study, we focus on the seeded batch cooling scenario. Following the framework introduced by \citet{Ramkrishna2000PopulationEngineering} and parameterized by \citet{Li2017ModelingCrystallization}, we use a one-dimensional PBM to describe the system \cite{Ramkrishna2014PopulationProspects, Randolph2012TheoryCrystallization}. The PBM describes the evolution of the crystal number density $n$ with respect to time $t$ and the characteristic crystal size $L$ as
\begin{equation} \label{eqn:pbe}
    \frac{\partial n}{\partial t} + G\frac{\partial n}{\partial L} = 0
\end{equation}
where $G$ represents the crystal growth rate. The method of moments \cite{Bowman2005Estimation:Moments, Pearson1936MethodLikelihood} is used to solve this partial differential equation by breaking it into a set of ordinary differential equations (ODEs) as implemented by \citet{Kim2023ModelingStrategy}. In the method of moments, the $i$th moment $\mu_i$ is defined as 
\begin{equation} \label{eqn:moment}
    \mu_i=\int^\infty_0 L^in(L)dL, \qquad i = 0,1,2,3, ...
\end{equation}
The zeroth through the third moment, $\mu_0$, $\mu_1$, $\mu_2$, $\mu_3$, are each proportional to the total number, length, surface area, and volume of crystals. Assuming a size-independent growth rate $G$ \cite{Li2017ModelingCrystallization, Worlitschek2004Model-BasedParacetamol}, the time derivative of each moment can be derived based on Eq. \eqref{eqn:pbe} and \eqref{eqn:moment} as
\begin{equation}
    \frac{\mathrm{d}\mu_i}{\mathrm{d}t} = iG\mu_{i-1}, \qquad i = 1, 2, 3, ...
\end{equation}
For the zeroth moment, which quantifies the total number of crystals, its time derivative can then be defined as the nucleation rate, which is
\begin{equation}
    \frac{\mathrm{d}\mu_0}{\mathrm{d}t} = B_1 + B_2
\end{equation}
where $B_1$ and $B_2$ represent the primary and secondary nucleation rates, respectively. Primary nucleation refers to the formation of crystals directly from a clear solution, while secondary nucleation describes new crystal formation facilitated by existing crystals \cite{Myerson2002HandbookCrystallization, Randolph2012TheoryCrystallization}. As shown in previous work, in seeded batch crystallization, the secondary nucleation rate is facilitated by the addition of seed crystals, and therefore primary nucleation is ignored in this case. Thus, the rate of change of the zeroth moment can be calculated solely by an empirical model of the secondary nucleation rate \cite{Omar2017CrystalReview}
\begin{equation}
\begin{split}
    \frac{\mathrm{d}\mu_0}{\mathrm{d}t} &= B_2 \\
    &= k_{b2}(S-1)^\alpha m_s^\beta
\end{split}
\end{equation}
where $k_{b2}$ is the secondary nucleation rate constant [(g crystals/kg solvent)$^{-\beta}$/(min$\cdot$g solvent)], $m_s$ is the crystal mass in a unit mass of solution [g crystals/kg solvent], $\alpha$ and $\beta$ are the model exponential parameters, and $S$ is the supersaturation. The supersaturation $S$ can be calculated from the concentration $C$ [g solute/g solvent] and the temperature-dependent saturated concentration $C_S$ [g solute/g solvent] as 
\begin{equation}
    S = C/C_S
\end{equation}
\begin{equation} \label{eqn:Cs}
    C_S = -16.17 + 1.765\times10^{-1}T - 6.439\times10^{-4}T^2 + 7.915\times10^{-7}T^3
\end{equation}
where $T$ is the process temperature in Kelvin, and the saturated concentration polynomial parameters were taken from previous work by \citet{Li2014ApplicationCrystallization}.

Given that the third moment $\mu_3$ corresponds to the volume of the crystals, the solute concentration $C$ in a batch process can be calculated from the mass balance as
\begin{equation} \label{eqn:C}
    C(t) = C_0 - k_{\nu} \rho \mu_3(t)
\end{equation}
with the time derivative
\begin{equation}
    \frac{\mathrm{d}C}{\mathrm{d}t} = - k_{\nu} \rho \frac{\mathrm{d}\mu_3}{\mathrm{d}t} = - 3k_{\nu} \rho G \mu_2
\end{equation}
where $C_0$ is the initial concentration [g solute/g solvent], $k_{\nu}$ is the volume shape factor of the crystals, and $\rho$ is the solid density of the crystals [g/cm$^3$]. The seed mass is intentionally set to be small, so it can be ignored from the mass balance. As mentioned above, under the assumption of size-independent growth, $G$ can be written as an Arrhenius-type equation with the absolute supersaturation driving force $C - C_S$ as
\begin{equation}
    G = k_g \mathrm{exp} \left( -\frac{E_{a_g}}{RT} \right) (C-C_S)^{\gamma_g}
\end{equation}
where $k_g$ is the pre-exponential factor for crystal growth [($\mu$m/min)(g solute/g solvent)$^{-\gamma_g}$], $E_{a_g}$ is the activation energies [J/mol], $\gamma_g$ is the exponential parameter of the driving force, and $R$ is the universal gas constant [J/(mol K)]. 

\subsection{Data Generation \& Preparation}
\label{subsec:data_gen}

Synthetic datasets are generated using the PBM and associated moment equations described in Section \ref{subsec:pbm}. The simulation setup is designed to mimic batch cooling crystallization experiments under dynamic temperature profiles. Following the seeded batch stage-cooling profile described by \citet{Li2017ModelingCrystallization}, each experiment consists of three stages: an initial isothermal plateau, a linear cooling phase, and a final isothermal hold period. The temperature evolution is constructed using the function:
\begin{equation} \label{eqn:temp}
    T(t) = 
    \begin{cases}
        T_{\text{plat}}, & 0\leq t < t_{\text{plat}}, \\
        T_{\text{plat}} + r(t-t_{\text{plat}}), & t_{\text{plat}} \leq t < t_{\text{cool}}, \\
        T_{\text{final}}, & t_{\text{cool}} \leq t < t_{\text{end}},
    \end{cases}
\end{equation}
where $T_{\text{plat}}$ is the initial plateau temperature [$\unit{\celsius}$], $r$ is the cooling rate [$\unit{\celsius}$/min], $t_{\text{plat}}$ is the plateau duration [min], $t_{\text{cool}}$ is the cooling time [min], $T_{\text{final}}$ is the final holding temperature at 0$\unit{\celsius}$ and $t_{\text{end}}$ is the end time of the experiment, which is set to 500 minutes here. Parameters $T_{\text{plat}}$, $r$, and $t_{\text{plat}}$ are randomly sampled from a uniform distribution within predefined ranges to produce a diverse set of experimental conditions, and $t_{\text{cool}}$ is calculated based on the difference between $T_{\text{final}}$ and $T_{\text{plat}}$. The initial solute concentration $C_0$ is also randomly varied from a uniform distribution to introduce variability in supersaturation dynamics. The synthetic temperature profiles and distribution of $T_{\text{plat}}$, $r$, $t_{\text{plat}}$, and $C_0$ can be found as Figure \ref{fig:temp_profile} to \ref{fig:dist_plat} in the SI.

For each simulated experiment, the ODE system in Section \ref{subsec:pbm} is solved using the LSODA method implemented in \textsc{SciPy} \cite{Virtanen2020SciPyPython} based on the generated temperature profile and the parameter values reported by \citet{Li2017ModelingCrystallization}. Five state variables are included in the forward solution, from $\mu_0$ to $\mu_3$ and $C$, and initial conditions for these five state variables are specified when solving the ODE system. A minimal value is intentionally set as the initial condition for $\mu_0$ to ensure the seeded condition of the simulated system, and the value is at least 3 orders of magnitude smaller than the numerical value of $\mu_0$ to make it negligible in the mass balance. Simulation details are available in Section \ref{subsec:SI_data} in the SI. Each simulation produces a time-series output for the concentration and the zeroth to the third moment. One thing to note is that moments are usually not directly measurable in experiments. Previous studies have already correlated FBRM signals with these moments, and Kim proposed a shallow NN framework to accurately convert FBRM measurements into moments data \cite{Kim2021DevelopmentApproaches, Lima2024NeuralCrystallization, OCiardha2012SimultaneousCrystallization}. Alternatively, an in-situ camera may be used to estimate moments. In this study, we assume a ``perfect sensor'' where moments and concentration are all directly accessible, and in our baseline case we assume that solubility is also perfectly calibrated.

Although noise-free simulations provide an ideal baseline for comparison, real experimental data contains measurement noise. To replicate this non-ideality in a controlled manner, Gaussian noise is introduced to both the moments and concentration trajectories in Section \ref{sec:noise}. The noise amplitude is set to 10\%, 30\%, and 100\% of the standard deviation of all values over each state trajectory (i.e. the standard deviation of all concentrations/moments from $t=0$ to $t=500$ min) to represent ``low-noise,'' ``mid-noise,'' and ``high-noise'' cases. The generated data profiles are provided in Figure \ref{fig:data_profile_noiseless} to \ref{fig:data_profile_10std} in the SI.

In addition to introducing controlled levels of noise to reflect the aleatoric uncertainty, adjustments are made to the solubility expression used within the PBM simulations to replicate the non-idealities associated with real measurements in Section \ref{sec:solubility}. Solubility values can be different because of the material purity and the analysis methods. Shifts are commonly reported in the crystallization field, where the differences between predicted and measured equilibrium solubility can reach several percent depending on the quality of the dataset and the fitting method used \cite{Granberg1999SolubilitySolvents, Hahnenkamp2010MeasurementIngredients, Kim2023ModelingStrategy, Widenski2010ComparisonCrystallization, Worlitschek2004Model-BasedParacetamol}. Introducing this systematic bias reflects the mismatch between ``true'' thermodynamic solubility and the working correlations typically used to correlate process simulations and experiments. To incorporate this, the saturated concentration, $C_{S}$, calculated as a polynomial function of temperature using Eq. \eqref{eqn:Cs} is uniformly shifted by 10\% throughout the temperature range in the constant solubility shift case. This bias serves as a source of epistemic model uncertainty that arises from imperfect knowledge of the system and the model.

We also consider the sampling frequency as another source of epistemic uncertainty in Section \ref{sec:sampling}. To mimic experimental constraints, in this case, only 10 runs are included in the training set, and we strategically evaluate cases that utilize 2, 3, 5 and 9 points per run, which correspond to 20, 30, 50 and 90 training points in total. We follow a non-uniform and approximately logarithmic sampling scheme. In the 2-point case, only the initial and final values of the state variables are available; in the 3-point case, one data point is added before $t_\text{plat}$ since the dynamics are faster, or have steeper gradient, in the early phase; in the 5-point case, three additional points are added at the three cooling stages, respectively, as shown in Eq. \eqref{eqn:temp}, and in the 9-point case, a roughly logarithmic spacing is used to create a denser early and sparser later sampling. The detailed sample points can be found in Section \ref{subsec:SI_downsamp} in the SI. One thing to note is that although the state variables are heavily down-sampled from one minute resolution, the complete temperature profile in minute resolution is assumed to be available since temperature is the controlled variable in this crystallization system. Integrating these sources of error into the synthetic dataset thus provides an additional layer of realism and allows an assessment of the robustness of different modeling strategies under conditions where key thermodynamic knowledge is not perfectly known.

To provide an unbiased evaluation of the model, the generated data is split into training, validation, and testing set in a 6:2:2 ratio, where in our case 100 synthetic runs are generated in total, leading to 60-20-20 training, validation, and testing runs. The training data are used to optimize the machine learning model, including network weights, biases, PBM parameters, and other learnable parameters. Here, the validation data are not used to optimize or tune model configuration hyperparameters (e.g. network width/depth, learning rate, etc.) but used only to select the ``best performing'' model via early stopping, which is a regularization technique that improves model generalization in unseen data \cite{Girosi1995RegularizationArchitectures, Yao2007OnLearning}. This procedure provides a relatively fair way to choose a single best model to evaluate model performance and prevent possible overfitting to the training data. Then, the testing data are used to provide a final unbiased evaluation of the model performance. In this paper, if not specified, the validation and test error all refer to the error based on the best performing model selected using early stopping based on the validation set. Before training, all state variables are normalized by their respective maximum values throughout the training set to ensure numerical stability and balance the contributions of each variable to the loss function. 

\subsection{Physics-informed Recurrent Neural Networks (PIRNNs)}

\begin{figure*}[h!]
    \centering\includegraphics[keepaspectratio=true,scale=0.45]{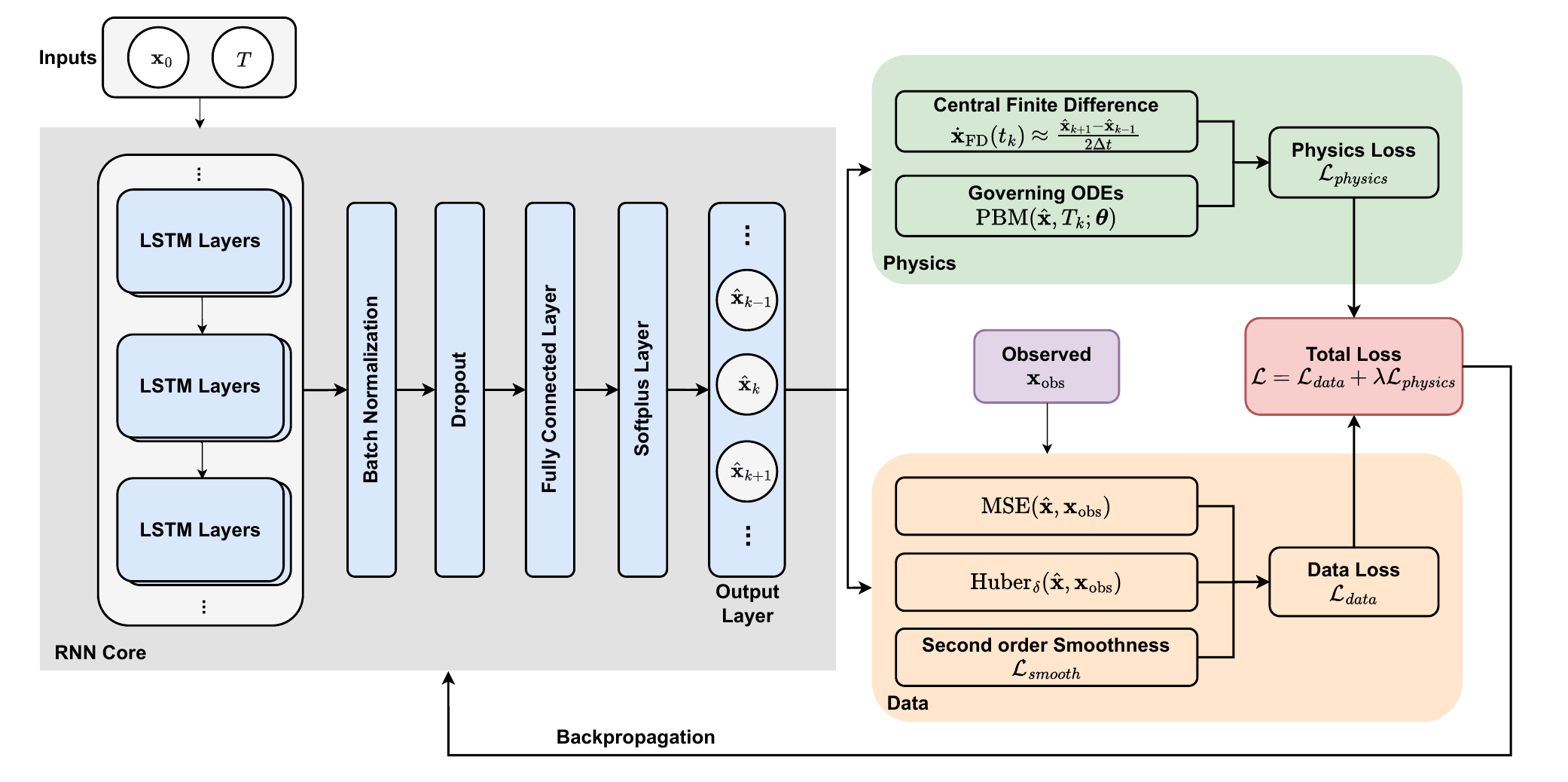}
    \caption{The PIRNN architecture and the loss terms.}
    \label{fig:pirnn_arch}
\end{figure*}

Following the principle of physics-informed neural networks (PINNs) introduced by \citet{Raissi2019Physics-informedEquations}, and building directly based on  the architecture of PIRNNs recently adapted for crystallization processes by \citet{Wu2023Physics-informedProcess}, we employ a hybrid framework that combines the sequential modeling ability of Long Short-Term Memory networks with physics-based constraints to model crystallization dynamics \cite{Hochreiter1997LongMemory}, as shown in Figure \ref{fig:pirnn_arch}. The PIRNN model consists of an LSTM encoder that processes concatenated state variables $\mathbf{x}(t)$ and control inputs $T(t)$, where
\begin{equation}
    \mathbf{x}(t) = \left[ \mu_0, \mu_1, \mu_2, \mu_3, C \right]^\intercal
\end{equation}
followed by batch normalization and dropout layers to prevent covariate shift and overfitting and improve generalization, and a fully connected linear decoder that outputs the predicted state variables at each forward step \cite{Hinton2012ImprovingDetectors, Ioffe2015BatchShift, Wan2013RegularizationDropconnect, Srivastava2014Dropout:Overfitting}. Two LSTM layers are used here to keep the architecture simple. Softplus activation is applied to the outputs to ensure nonnegative predictions. The detailed design and hyperparameters (i.e, number of nodes/layers, dropout rate, etc.) can be found in Table \ref{tab:pirnn_hyperparameters} in the SI. During training, the PIRNNs take an initial state $\mathbf{x}_0 \in \mathbb{R}^5$ and temperature control series $T \in \mathbb{R}^{N \times 1}$ for $N$ timestamps and provide a predicted trajectory $\hat{\mathbf{x}} \in \mathbb{R}^{N \times 5}$ for the five state variables.

The total training loss $\mathcal{L}$ is a weighted sum of data and physical loss terms:
\begin{equation} \label{eqn:L_total}
    \mathcal{L} = \mathcal{L}_{data} + \lambda \mathcal{L}_{physics}
\end{equation}
where $\lambda$ is a weight hyperparameter that governs the relative importance of data fitting and physics matching. Data loss $\mathcal{L}_{data}$ consists of three main parts: mean squared error (MSE) between the predicted states and observed training states, Huber loss, and smoothness loss, where
\begin{equation}
    \mathcal{L}_{data} = \omega \text{MSE}(\hat{\mathbf{x}}, \mathbf{x}_\text{obs}) + (1-\omega)\text{Huber}_{\delta}(\hat{\mathbf{x}}, \mathbf{x}_\text{obs}) + \mathcal{L}_{smooth}
\end{equation}
\begin{equation}
    \omega = \sigma(-\eta_{noise})
\end{equation}
\begin{equation}
    \text{MSE}(\hat{\mathbf{x}}, \mathbf{x}_\text{obs}) = \frac{1}{N}\sum^{N}_{k=1} \left\Vert \hat{\mathbf{x}}_{k} - \mathbf{x}_{k,\text{obs}} \right\Vert^2_2
\end{equation}
\begin{equation} \label{eqn:L_smooth}
    \mathcal{L}_{smooth} = \frac{1}{N-2} \sum^{N-1}_{k=2} \left\Vert \hat{\mathbf{x}}_{k+1} - 2\hat{\mathbf{x}}_{k} + \hat{\mathbf{x}}_{k-1} \right\Vert ^2
\end{equation}
Here, $\omega$ is a sigmoid function with respect to a learnable noise scale parameter $\eta_{noise}$ that allows adaptive mixing of MSE and Huber loss. This approach is inspired by the idea of heteroscedastic likelihood loss that allows the model to internally assign a lower weight to data points that are believed to be noisier \cite{Kendall2017WhatVision}. Huber loss is a robust regression loss that combines the sensitivity of MSE for small residuals with the robustness of mean absolute error (MAE) for large residuals using a transition threshold $\delta$, specifically,
\begin{equation}
    \text{Huber}_{\delta}(\hat{\mathbf{x}}, \mathbf{x}_\text{obs}) = 
    \begin{cases}
        0.5(\hat{\mathbf{x}} - \mathbf{x}_\text{obs})^2, & \text{if} |\hat{\mathbf{x}} - \mathbf{x}_\text{obs}| <\delta\\
        \delta(|\hat{\mathbf{x}} - \mathbf{x}_\text{obs}| - 0.5\delta) & \text{otherwise}
    \end{cases}
\end{equation}
Here, if the errors are smaller than the threshold, the loss is quadratic, which is identical to MSE, which ensures smooth gradients and strong penalization; on the other hand, for large errors, the loss grows linearly with the residual, similar to MAE, which helps reduce the influence of outliers and noisy spikes, preventing them from dominating the training process. The combination also ensures that a smooth gradient that exists everywhere, unlike MAE which is nondifferentiable when the residual is zero, making Huber loss an ideal choice here for gradient-based optimization \cite{Huber1964RobustParameter}. Smoothness loss $\mathcal{L}_{smooth}$ defined in Eq. \ref{eqn:L_smooth} is also used to obtain smoother temporal behavior and to further prevent noise overfitting. It penalizes rapid oscillations in predicted states using a second-order finite difference, inspired by regularization techniques in signal processing and optimal control theory. The relative weights of these loss terms are arbitrarily set to be equal to each other.

Following typical PINN design, the physics loss $\mathcal{L}_{physics}$ is included in Eq. \ref{eqn:L_total} to enforce consistency between the learned trajectories and the underlying physics \cite{Raissi2019Physics-informedEquations}. It penalizes the residual between a finite difference approximation to the time derivative based on the predicted state
\begin{equation}
    \dot{\mathbf{x}}_\text{FD}(t_k) \approx \frac{\hat{\mathbf{x}}_{k+1} - \hat{\mathbf{x}}_{k-1}}{2 \Delta t}, \quad k = 2, ...,N-1
\end{equation}
and the underlying rates of change defined in the PBM moment ODEs using an MSE criterion as
\begin{equation}
    \mathcal{L}_{physics} = \text{MSE}\left( \dot{\mathbf{x}}_\text{FD}, \text{PBM}(\hat{\mathbf{x}}, T_k; \boldsymbol{\theta}) \right)
\end{equation}
where $\boldsymbol{\theta}$ contains all the learnable parameters of the PIRNNs, including weights and bias, and the kinetic parameters of the PBM \cite{Lim2022PhysicsMethod}. In this paper, all PBM parameters in all conducted case studies are initialized to 1 to avoid biasing our model from known knowledge. The PBM parameters learned by PIRNNs can then be compared with the reference values provided in Table \ref{tab:pirnn_param}.

In this paper, the model performance is mainly evaluated through two metrics: the PIRNN prediction and the prediction of the re-evaluated ODE. The PIRNN prediction refers to the direct forward prediction using weights and bias of the network, and the re-evaluated ODE means solving the ODE system in Section \ref{subsec:pbm} using the PBM parameters estimated from the PIRNN optimization.

\section{Results and Discussion} 
\label{sec:results}

The performance of the PIRNN was evaluated under varying levels of measurement noise and training data availability. Aleatoric signal noise and epistemic uncertainties in the forms of solubility mis-calibration and reduced sampling frequencies were introduced in a controlled manner and independently investigated.

\subsection{Signal Noise} \label{sec:noise}

\begin{figure*}[h!]
    \centering\includegraphics[keepaspectratio=true,scale=0.45]{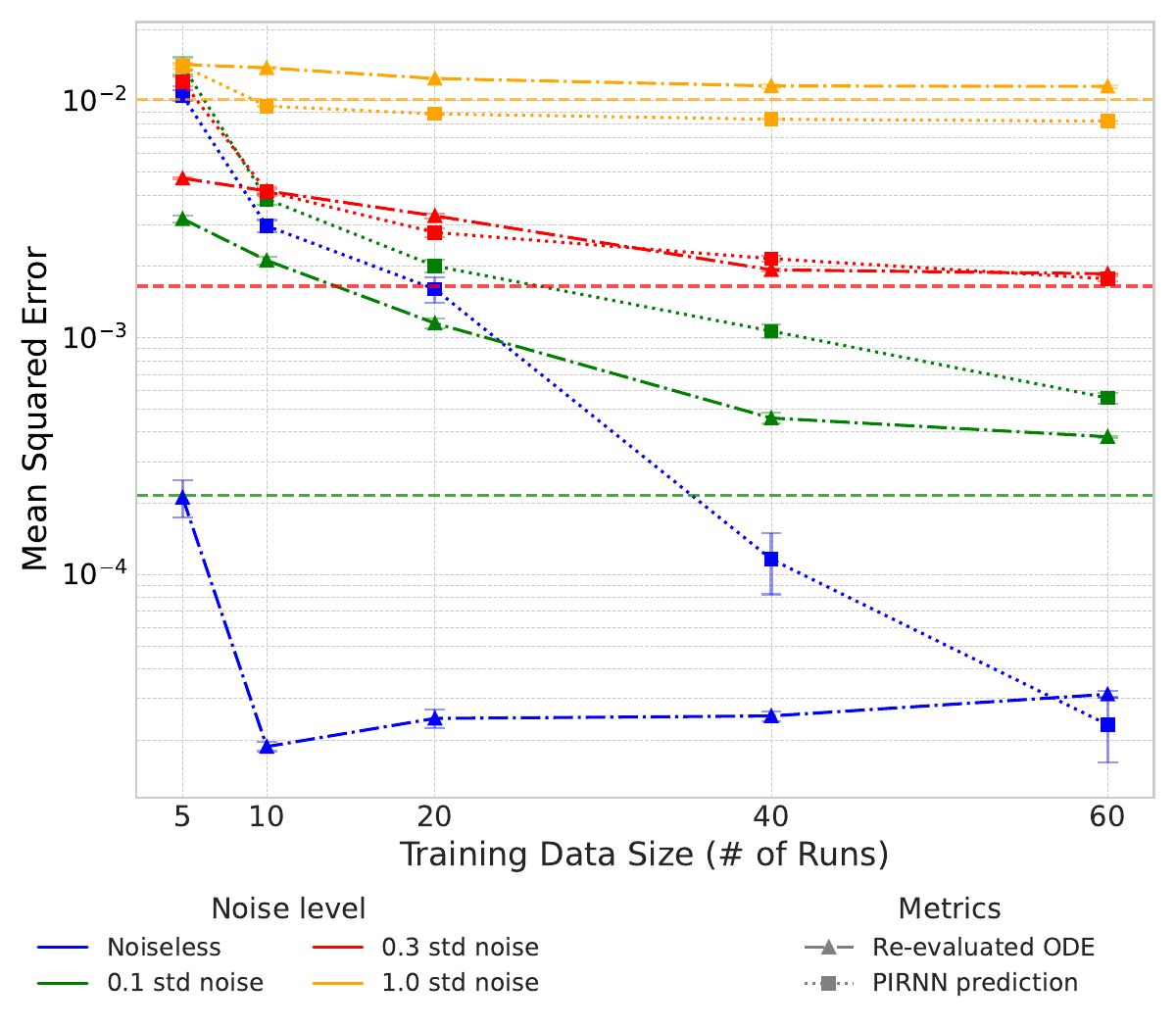}
    \caption{Testing data prediction MSE as a function of training data under varying noise levels in two different metrics: the PIRNN forward prediction and re-evaluated ODE calculated using kinetic parameters learned from the PIRNN. The MSE is shown for noiseless, 0.1 std, 0.3 std, and 1.0 std noise conditions in increasing numbers of training runs. The horizontal dashed lines indicate the reference error between the noisy data and the ground-truth noiseless data.}
    \label{fig:model_perform}
\end{figure*}

Figure \ref{fig:model_perform} shows the PIRNN testing loss and physical consistency across varying levels of noise. As expected, increasing the training data size generally reduces the testing MSE, indicating an improved generalization of the model as more data are included. In noiseless regimes, both evaluation metrics achieve a low error level below $10^{-4}$. However, the direct PIRNN forward prediction shows a stronger dependency on the training data size, while the re-evaluated ODE curve using the PIRNN estimated parameters exhibits a stable behavior when using more than 10 training runs. It suggests that, under ideal conditions, the PIRNN model is able to extract reliable kinetic parameters from just a small amount of data. When a large amount of data is available (60 training runs), the flexible PIRNN structure can interpolate the state dynamics more precisely than the rigid ODE integration, which remains sensitive to slight deviation in the estimated parameter, indicating a possible limitation of using finite difference. Switching to a more exact derivative calculation method might solve this problem, but detailed evaluation of the differential method is beyond the scope of this paper. This result suggests a possible dual usage of the PIRNN model: it can be used for both direct state prediction and parameter extraction. 

When noise is introduced, both the PIRNN prediction and the re-evaluated ODE errors plateau around at a higher magnitude, which is close to the reference error between the noisy data and ground-truth noiseless data that is shown as horizontal dashed line in Figure \ref{fig:model_perform}. The plateau behavior around these horizontal lines suggests that the error elevation is not due to the performance degradation of the model itself, but due to the irreducible aleatoric uncertainty inherent in the data because of the addition of noise. At low noise level (0.1 std), the re-evaluated ODE error is generally lower than the direct PIRNN prediction, showing that low level error would not have significant effects on the quality of extracted physics. However, at high noise level (1.0 std), the re-evaluated ODE shows a higher error level than the PIRNN forward prediction, suggesting a degradation in the recovered physics. One thing to note is that at high noise level (1.0 std), the PIRNN prediction error is lower than the reference level, indicating a possible overfitting of the noisy data. This might be solved by tweaking the relative weights of different loss terms, but for a controlled comparison consideration, the weights of different loss terms stay unchanged across different noise level case studies. Overall, in contrast to the ideal case where the re-evaluated ODE remains stable, here it clearly scales with the input noise and training data size, demonstrating that the accuracy of the recovered physics is ultimately bound by the quality of the observed data. 

\begin{figure*}[h]
    \centering\includegraphics[keepaspectratio=true,scale=0.6]{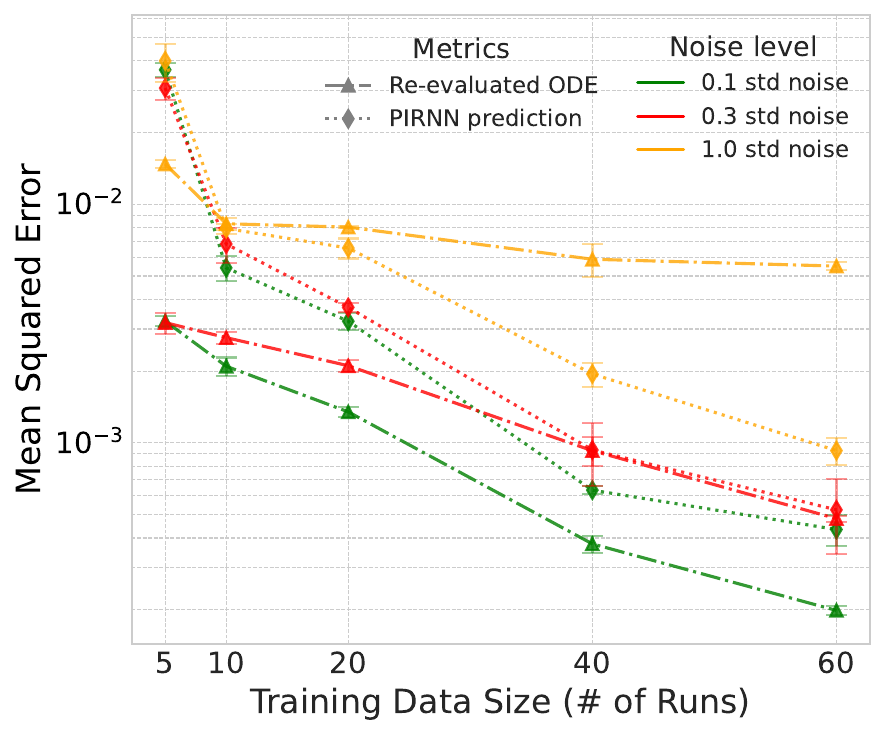}
    \caption{Testing MSE between the PIRNN predicted states and learned PBM parameters and the ground-truth noisefree reference at different noise levels.}
    \label{fig:noiseless_mse}
\end{figure*}

Figure \ref{fig:noiseless_mse} shows the testing errors calculated based on noiseless data instead of noisy data. The testing errors calculated with noiseless data decrease to around $10^{-3}$ and show similar behavior at different noise levels. This further confirms that the error elevation in Figure \ref{fig:model_perform} is caused by intrinsic information loss due to the addition of noise, but not the degradation of the performance of the model, and the presence of noise has minimal effects on the generalization and prediction capability of the model. The re-evaluated ODE error for high-noise case remains around an order of magnitude higher than that of the low-noise case even at large training size (60 runs). This indicates that while the physics constraints alleviate overfitting, high-magnitude aleatoric uncertainty still inevitably introduces a residual bias in the parameter estimation that prevents the convergence of physics to the ground-truth reference state. The two evaluation metrics also show a switching behavior as noise level increases. In the low-noise case, the re-evaluated ODE error is lower than the PIRNN prediction; in the mid-noise case, both metrics converge into a similar error level, and in the high-noise case, the PIRNN prediction clearly outperforms the re-evaluated ODE. This behavior suggests that the dual usage of PIRNNs should be wisely decided based on the quality of the available data.

\begin{figure*}[h]
    \centering\includegraphics[keepaspectratio=true,scale=0.54]{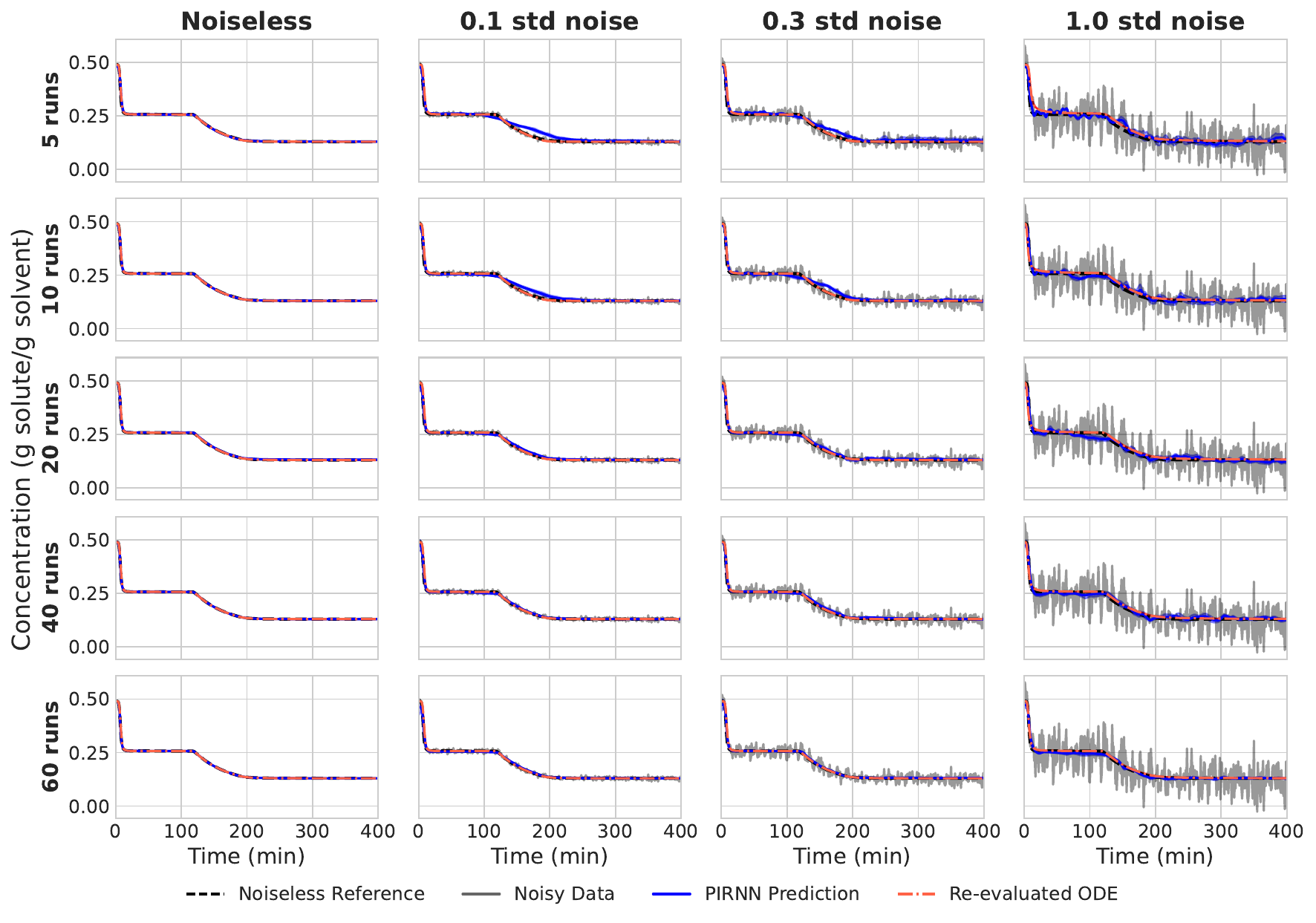}
    \caption{Concentration prediction of a set of test data using 10, 20, 40, and 60 training runs. The 95\% confidence intervals are provided. The original noisy data and ground-truth noiseless reference are shown for each case. Prediction of other state variables can be found as Figure \ref{fig:test_pred_mu0} to \ref{fig:test_pred_mu3} in the SI.}
    \label{fig:test_pred}
\end{figure*}

An example of a randomly selected testing concentration trajectory and the results predicted with models trained using different amounts of training runs at different noise levels are shown in Figure \ref{fig:test_pred}. Across all cases, both the PIRNN prediction and the re-evaluated ODE curves follow the ground-truth reference. For noiseless to mid-noise cases (noiseless, 0.1 std, and 0.3 std noise), the predicted concentration profiles completely overlap with the reference, indicating a precise and accurate reconstruction of the crystallization dynamics. As the noise level increases from 0.3 to 1.0 std, the PIRNN prediction starts to show slight noise overfitting, but the prediction is still overall smooth and the fluctuation is minimal, demonstrating the high noise tolerance of the PIRNN structure \cite{Nai2024Micro-kineticNetworks}. Meanwhile, at 1.0 std noise, the re-evaluated ODE curve is consistent with all other cases, suggesting the robustness and the physical consistency of PIRNNs even at substantial amount of noise. Based on these results, the PIRNN architecture effectively constrains the model to follow physically plausible trajectories. Even under high aleatoric uncertainty, the model captures both the magnitude and timing of important dynamic features, such as the supersaturation decay and the final steady state. The re-evaluated ODE curves, obtained by integrating the learned kinetic parameters, further confirm the consistency between learned dynamics and the governing PBM.

Another way of evaluating the performance of the PIRNN is to analyze the retrieved parameter values of the physical model. Table \ref{tab:pirnn_param} compares the kinetic parameters estimated by the PIRNN with the reference values previously reported by \citet{Li2017ModelingCrystallization}, which are used to generate the synthetic data. In the noiseless case, all learned parameters converge rapidly toward the literature references, with deviation below 10\% for all quantities at 10 or more runs. The standard deviations are generally small and decrease as more runs are involved in the training. The exception is the case of $k_{b2}$ and $\beta$, which occurs because these are highly correlated parameters that appear in the same term of the model. As seen in Section \ref{subsec:pbm}, the PBM parameters are not solely independent, especially when the available amount of data is limited. Therefore, the decrease in $k_{b2}$ is compensated by the increase in $\beta$, as the available amount of data is not sufficient to deconvolute the effects of the two parameters.

As the aleatoric uncertainty level increases, the precision of the estimated PBM parameters naturally degrades, as reflected by the increasing standard deviation and deviations from reference values. However, comparing these parameter discrepancies with the re-evaluated ODE curve in Figure \ref{fig:test_pred}, it is clear that this parameter accuracy degradation does not lead to a substantial and proportional failure in dynamic prediction. This decoupling suggests that the inverse problem becomes practically unidentifiable for a unique solution under the presence of high aleatoric uncertainty. In this case, the optimization landscape flattens and allows highly correlated parameters to compensate each other. Therefore, the PIRNN starts to extract a kinetically equivalent parameter set that reproduces similar state trajectories. This behavior further suggests the robustness of the dual usage of PIRNNs. Although high measurement noise limits the uniqueness of the intrinsic kinetic parameters, the framework still effectively acts as a robust state estimator that can accurately reconstruct the process dynamics.

\begin{table}[h]
\centering
\renewcommand{\arraystretch}{1.2}
\begin{adjustbox}{max width=\textwidth}
\begin{tabular}{lcccccc}
\toprule
\textbf{PBM Parameter} & $k_{b2}$ & $\alpha$ & $\beta$ & $k_{g}$ & $E_{ag}$ & $\gamma_{g}$ \\
\midrule
\textbf{Reference Value} & $6.000\times10^{3}$ & 2.080 & 0.713 & $2.730\times10^{5}$ & $4.130\times10^{4}$ & 1.240 \\
\midrule
\multicolumn{7}{l}{\textbf{Noiseless}} \\
\hspace{1em}5 runs  & $5.455(43)\times10^{3} $ & 2.041(8) & 0.725(1) & $8.626(7044)\times10^{4}$ & $3.792(161)\times10^{4}$ & 1.239(6) \\
\hspace{1em}10 runs & $5.467(40)\times10^{3}$ & 2.052(4) & 0.726(1) & $2.036(266)\times10^{5}$ & $4.066(34)\times10^{4}$ & 1.228(1) \\
\hspace{1em}20 runs & $5.400(30)\times10^{3}$ & 2.054(2) & 0.729(1) & $2.213(316)\times10^{5}$ & $4.088(40)\times10^{4}$ & 1.227(1) \\
\hspace{1em}40 runs & $5.380(10)\times10^{3}$ & 2.057(1) & 0.729(1) & $2.465(108)\times10^{5}$ & $4.120(12)\times10^{4}$ & 1.225(1) \\
\hspace{1em}60 runs & $5.376(9)\times10^{3}$ & 2.058(0) & 0.730(0) & $2.563(65)\times10^{5}$ & $4.130(7)\times10^{4}$ & 1.225(1) \\
\midrule
\multicolumn{7}{l}{\textbf{0.1 std Noise}} \\
\hspace{1em}5 runs  & $5.184(49)\times10^{3} $ & 1.878(5) & 0.683(3) & $1.259(984)\times10^{2}$ & $2.116(191)\times10^{4}$ & 1.279(9) \\
\hspace{1em}10 runs & $4.839(61)\times10^{3}$ & 1.944(5) & 0.712(4) & $2.083(1072)\times10^{3}$ & $2.875(150)\times10^{4}$ & 1.242(7) \\
\hspace{1em}20 runs & $5.220(55)\times10^{3}$ & 1.961(8) & 0.697(3) & $1.169(352)\times10^{4}$ & $3.345(81)\times10^{4}$ & 1.232(9) \\
\hspace{1em}40 runs & $4.865(78)\times10^{3}$ & 1.975(3) & 0.714(4) & $7.437(1771)\times10^{4}$ & $3.839(71)\times10^{4}$ & 1.204(4) \\
\hspace{1em}60 runs & $4.791(42)\times10^{3}$ & 1.974(3) & 0.718(2) & $1.257(179)\times10^{5}$ & $3.982(41)\times10^{4}$ & 1.198(1) \\
\midrule
\multicolumn{7}{l}{\textbf{0.3 std Noise}} \\
\hspace{1em}5 runs  & $4.570(286)\times10^{3} $ & 2.065(56) & 0.717(20) & $1.450(731)\times10^{1}$ & $1.527(130)\times10^{4}$ & 1.434(13) \\
\hspace{1em}10 runs & $4.884(177)\times10^{3}$ & 2.104(36) & 0.716(9) & $2.200(862)\times10^{2}$ & $2.249(113)\times10^{4}$ & 1.379(12) \\
\hspace{1em}20 runs & $6.272(300)\times10^{3}$ & 2.165(50) & 0.669(10) & $3.171(1737)\times10^{3}$ & $2.914(53)\times10^{4}$ & 1.368(11) \\
\hspace{1em}40 runs & $5.259(274)\times10^{3}$ & 2.069(33) & 0.704(15) & $3.517(1130)\times10^{4}$ & $3.595(107)\times10^{4}$ & 1.288(11) \\
\hspace{1em}60 runs & $4.984(228)\times10^{3}$ & 2.059(30) & 0.716(11) & $6.538(1755)\times10^{4}$ & $3.770(72)\times10^{4}$ & 1.270(8) \\
\midrule
\multicolumn{7}{l}{\textbf{1.0 std Noise}} \\
\hspace{1em}5 runs  & $2.245(454)\times10^{3} $ & 2.322(185) & 0.779(52) & $2.475(3385)\times10^{1}$ & $1.524(305)\times10^{4}$ & 1.629(42) \\
\hspace{1em}10 runs & $3.851(644)\times10^{3}$ & 2.495(235) & 0.688(40) & $1.517(1381)\times10^{2}$ & $2.022(308)\times10^{4}$ & 1.615(69) \\
\hspace{1em}20 runs & $5.527(894)\times10^{3}$ & 2.381(89) & 0.622(44) & $8.688(7578)\times10^{3}$ & $2.521(211)\times10^{4}$ & 1.573(39) \\
\hspace{1em}40 runs & $5.834(927)\times10^{3}$ & 2.601(296) & 0.620(41) & $1.059(874)\times10^{5}$ & $3.774(240)\times10^{4}$ & 1.496(66) \\
\hspace{1em}60 runs & $5.228(578)\times10^{3}$ & 2.540(268) & 0.648(29) & $1.474(931)\times10^{5}$ & $3.915(204)\times10^{4}$ & 1.446(81) \\
\bottomrule
\end{tabular}
\end{adjustbox}
\vspace{3pt}
\caption{The PIRNN estimated PBM parameters under different noise levels across different training data sizes. The reference value are from previous work by \citet{Li2017ModelingCrystallization}. All parameters are optimized in logarithm scale to guarantee positivity. Standard deviations are provided using parenthesis notation, where the number in parenthesis is the numerical uncertainty value referred to the corresponding last digits. For example, 1.974(3) = 1.974 $\pm$ 0.003. Details about this parenthesis notation can be found in reference \cite{Mohr2025CODATA2022}.}
\label{tab:pirnn_param}
\end{table}


\subsection{Solubility Shift} \label{sec:solubility}

A common scenario in real crystallization data is uncertainty in the calibration of the solubility model, which may drift between runs. Here, as described in Section \ref{subsec:data_gen}, we introduce a constant 10\% shift in the solubility relationship and evaluate the ability of the PIRNN to learn this deviation directly from the measured data. For simplicity, we have omitted signal noise from this analysis. 

To deal with this model mis-calibration, PIRNNs use physics regularization weights $\lambda$ as a tunable parameter that controls the competition between empirical flexibility and mechanistic understanding. At low $\lambda$ the model behaves like pure RNNs, prioritizing minimization of data residuals. Although it allows the network to stick with the observed dynamics despite the solubility shift, it sacrifices the useful information provided by other physical laws in the PBM, such as nucleation and growth rate. In contrast, a high $\lambda$ makes the PBM equations nearly hard constraints. It enforces strict physical consistency but forces the prediction to follow a biased solubility curve, ignoring the corrective information present in the experimental measurements. Therefore, identification of a reasonable regularization regime, where physics can guide the optimization toward physically plausible solutions without ignoring the observed data, becomes important in addressing the effect of epistemic uncertainty.

\begin{figure*}[h!]
    \centering
    \begin{subfigure}{0.49\textwidth}
    \centering\includegraphics[keepaspectratio=true,scale=0.5]{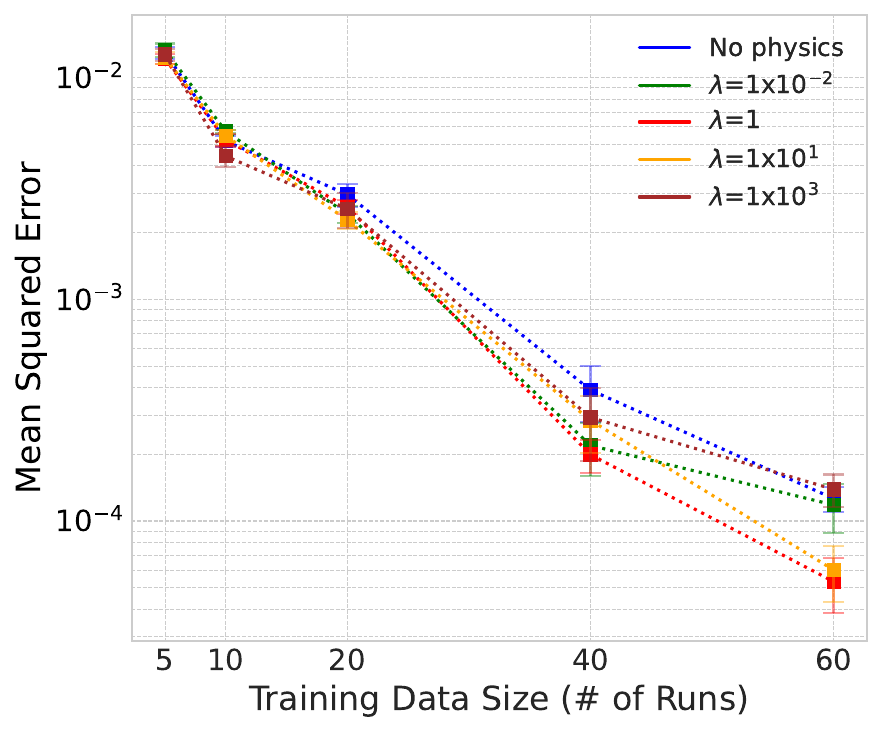}
    \caption{}
    \end{subfigure}
    \begin{subfigure}{0.49\textwidth}
    \centering\includegraphics[keepaspectratio=true,scale=0.5]{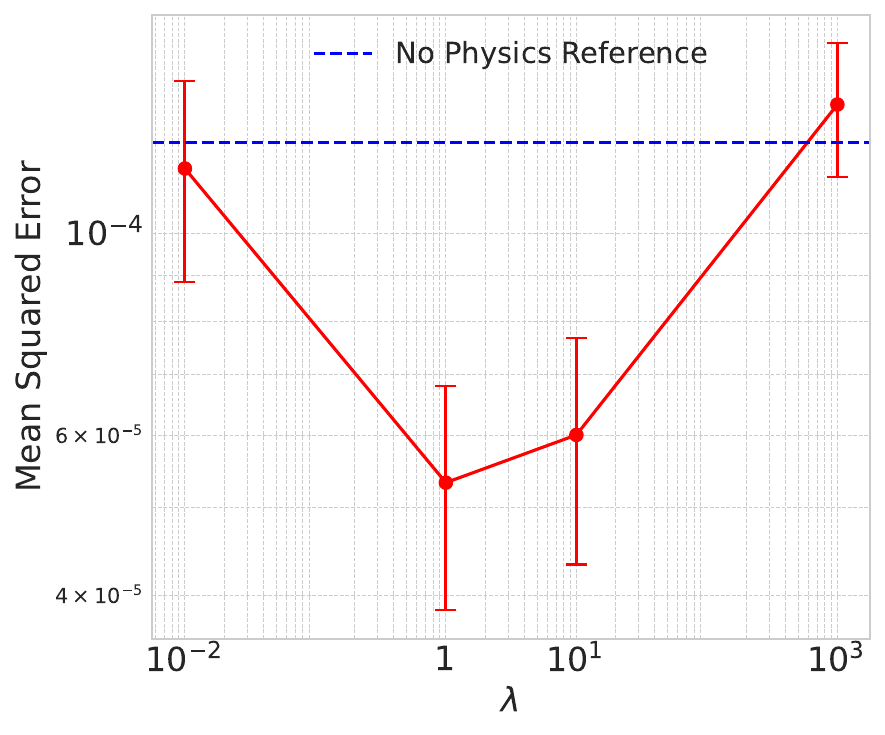}
    \caption{}
    \end{subfigure}
    \caption{Testing error comparison under shifted solubility model for varying strengths of physics regularization. (a) The MSE on test data is plotted as a function of training data size for different physics weight ($\lambda$ in Eq. \ref{eqn:L_total}) from $10^{-2}$ to $10^{3}$. (b) A closer look of the MSE as a function of $\lambda$ for 60 training runs case.}
    \label{fig:lambda_sweep}
\end{figure*}

Figure \ref{fig:lambda_sweep} illustrates the effect of varying the physics weighting coefficient on the test MSE under this model-mismatch condition. It can be observed that the inclusion of physics-based regularization significantly improves model generalization compared to the purely data-driven baseline or a weakly ($\lambda = 10^{-2}$) regularized comparison case. When the weight of physics is relatively low compared to data loss, the training is dominated by data, and the loss curve aligns closely with the purely data-driven baseline. For small datasets ($\leq$10 runs), all models show similar error levels, suggesting that the amount of data available is the main limitation. As training size increases, moderate physics ($\lambda = 1 \times 10^{[0,1]}$ ) weighting leads to faster convergence and orders-of-magnitude improvement in test performance, with the testing MSE decreasing below $10^{-4}$ at 60 runs, even though the embedded physics are not exactly matched with the training data. Here, the inclusion of physics as soft constraints shrinks the hypothesis space to functions that approximately obey the ``ground-truth'' physical model. Although there is a mismatch between the data and the model, the structure of the model imparts useful information about the smoothness and the correlations between the quantities. If the mismatch is moderate, the inclusion of physics regularization can still reduce the model variance. At the same time, because the physical laws are included as soft constraints, the network maintains sufficient flexibility to compensate for the solubility error without underfitting the observed dynamic trends. This balance follows the classical bias-variance tradeoff that is widely used in the machine learning field, where the bias introduced by the included physics is smaller than the variance it removes \cite{Belkin2019ReconcilingTrade-off, Geman1992NeuralDilemma, Girosi1995RegularizationArchitectures, Neal2018ANetworks, Poggio1987ComputationalTheory}. The approximated PBM serves as an inductive bias that regularizes the solution while allowing data-driven adaptation to epistemic uncertainty, leading to a higher generalization of the model and a more stable training process. 


\begin{figure*}[h!]
    \centering\includegraphics[keepaspectratio=true,scale=0.50]{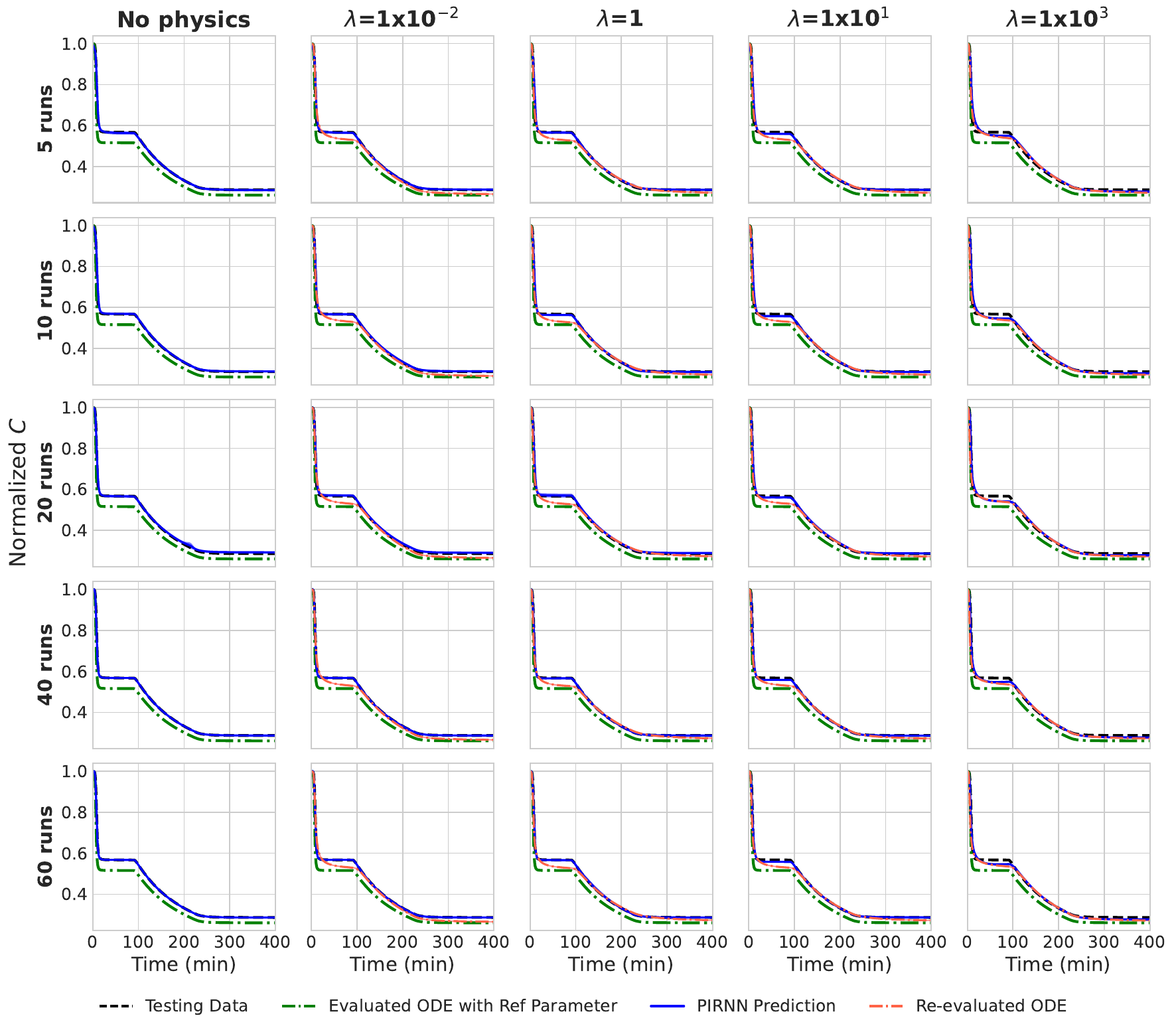}
    \caption{Normalized $C$ prediction of a set of test data using 5, 10, 20, 40, and 60 training runs at different $\lambda$ when 10\% constant solubility shift is applied. Re-evaluated ODE curves are not shown for no physics case since the PBM parameters are not included in optimization when physics loss is not included. The ODE curves are evaluated using two different sets of parameter: the reference values (green) and the PIRNN learned values (red). The prediction of other state variables are available in Figure \ref{fig:test_solushift_mu0} to \ref{fig:test_solushift_mu3} in the SI.}
    \label{fig:test_solushift_c}
\end{figure*}

Interestingly, excessive physics weighting reduces model performance, as indicated by visible error increases when the training size exceeds 40 runs or when the regularization strength exceeds $10^1$. This is because the network becomes biased towards the physics model that is based on an incorrect solubility model, limiting its ability to adapt to the empirical data space. This can be seen in Figure \ref{fig:test_solushift_c}, where both the PIRNN prediction and the re-evaluated ODE start to deviate from the testing data when $\lambda = 1 \times10^{3}$, although they still both perform better than direct ODE integration with the reference parameter values. This indicates an overly rigid adherence to potentially misspecified constraints. Here, compared to the moderate physics weighting case, the physics constraint acts as an inductive bias that should be optimally balanced with data fidelity, and the data and model loss start to compete with each other, leading to expected model performance degradation. Thus, in the presence of epistemic uncertainty, direct PIRNN prediction clearly outperforms ODE integration with reference parameter values, and with physics regularization the re-evaluated ODE is also closer to the testing data, even though it cannot be perfect because of the model mismatch.

Overall, the inclusion of appropriately weighted physics improves model robustness, enabling PIRNNs to compensate for epistemic uncertainty while maintaining a low generalization error. Importantly, these results highlight a distinct advantage of hybrid architecture, where direct PIRNN forward prediction delivers better accuracy compared to ODE evaluation using either reference or learned parameters, which is shown in Figure \ref{fig:test_solushift_c}. Because the underlying physics contains structural solubility bias, any strict integration of the governing equations propagates this error into the state trajectory. In contrast, PIRNNs act as flexible corrections, using physics for regularization while learning a trajectory that matches empirical observations. This result demonstrates that hybrid models are a promising candidate to bridge the gap between imperfect domain knowledge and real world observations.

\subsection{Sampling Frequency Limitations} \label{sec:sampling}

\begin{figure*}[h!]
    \centering\includegraphics[keepaspectratio=true,scale=0.6]{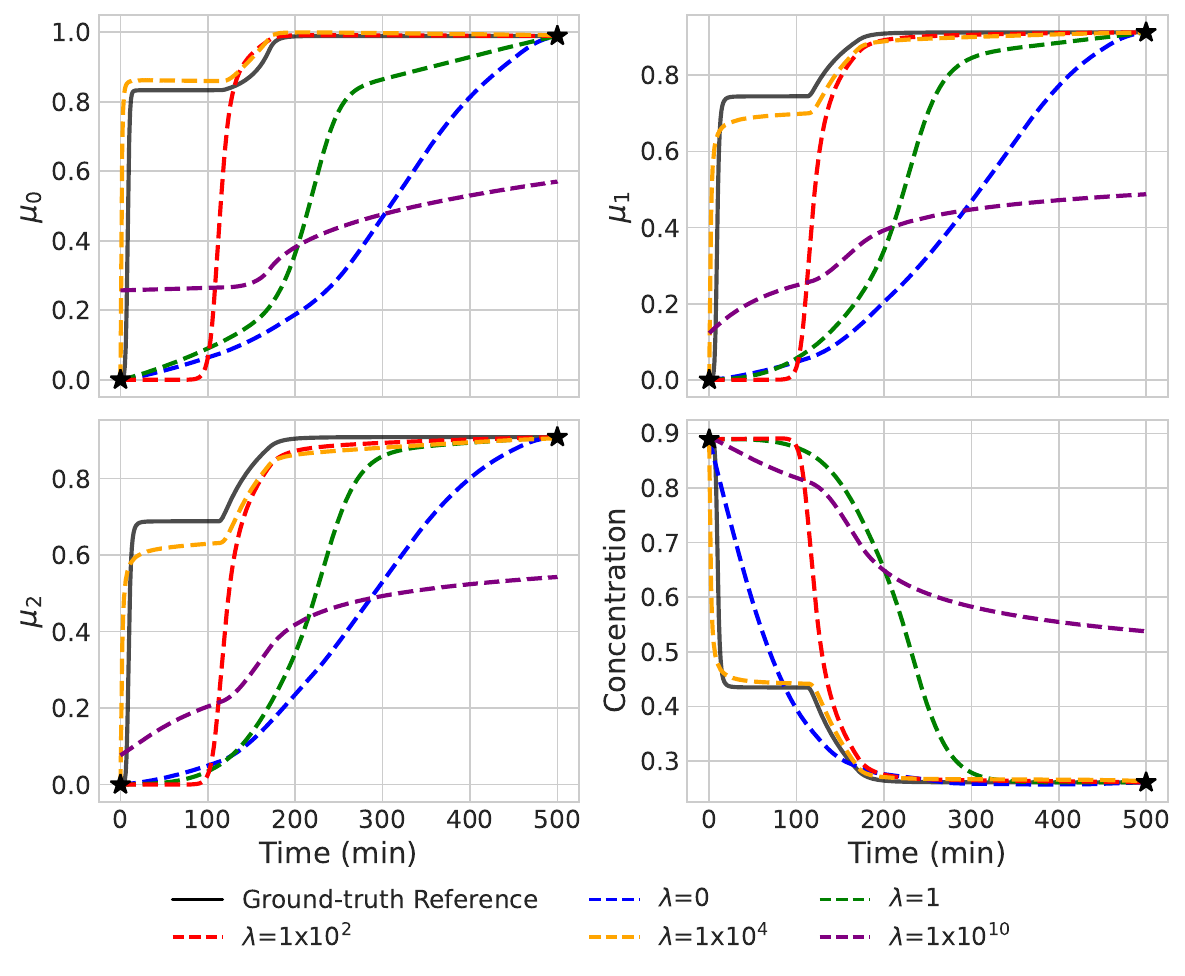}
    \caption{The PIRNN predicted trajectories for normalized $\mu_0$, $\mu_1$, $\mu_2$, and $C$ when only the initial and ending points are used for training (black stars in the figure). The ground-truth references are shown as black solid lines. $\mu_3$ is not shown here because it can be expressed via a mass balance using $C$.}
    \label{fig:downsamp_2}
\end{figure*}

Another common challenge in practical crystallization scenarios is the limitation on sampling frequency. To evaluate the robustness of PIRNNs under sampling uncertainty, i.e. highly sparse measurement conditions, we first considered an extreme scenario in which only the initial ($t$ = 0 min) and final ($t$ = 500 min) time points of the trajectory of each state variable were used during training. This setting mimics situations where PAT measurements are unavailable over most of the batch and only sieved crystal mass is accessible at the end of the run. Figure \ref{fig:downsamp_2} shows the normalized predicted trajectories for $\mu_0$, $\mu_1$, $\mu_2$, and $C$ under varying physics strengths. In the absence of physical constraints ($\lambda$ = 0), a purely data-driven model fails to capture the transient dynamics, producing essentially an arbitrary smooth interpolation between the start and end points that misrepresents the cooling-induced nucleation and growth events. In contrast, at the other extreme with excessive physics regularization ($\lambda$ = $10^{10}$), the model becomes almost purely ``physics-driven'', but it has no accurate prior knowledge of the physical parameters (i.e. it satisfies the differential equations, but the parameters of the differential equations are not consistent with the data). Since all PBM parameters are initialized to 1, which is far from the reference values, the optimizer can converge to a trivial solution by selecting physical parameters that make the PBM calculated derivatives decay to nearly zero. At the same time, because the relative weight on the data loss is negligible compared to the physics, the optimizer can choose to deviate from the known boundary points to reduce the finite-difference derivatives, therefore minimizing the physics loss term to $\sim$0. At intermediate regime, it can be seen that introducing physics constraints progressively, as $\lambda$ goes from 1 to $10^2$ to $10^4$, improves the model's ability to recover the unobserved dynamics. At the optimal point ($\lambda = 10^4$), the physics and data loss terms can be balanced, which allows the information from the boundary points to influence the learned PBM parameters, leading to physically meaningful trajectories. It is important to note that although the state measurements only include the boundary values, the model has access to the complete temperature trajectory since it is a controlled variables in experiments. This also allows the PIRNN to capture dynamics that are strongly coupled to temperature changes even when state measurements are not fully accessible.

\begin{figure*}[h!]
    \centering\includegraphics[keepaspectratio=true,scale=0.7]{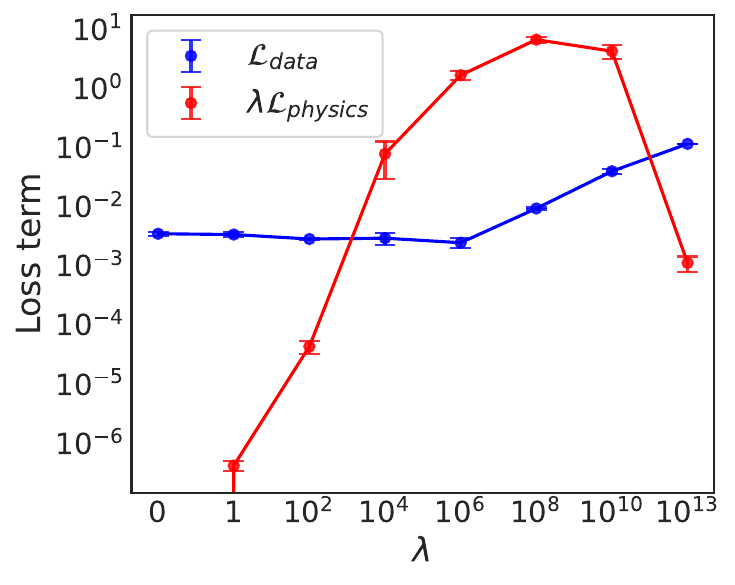}
    \caption{Data and PBM model training loss for different $\lambda$ values when only the initial and ending points are used for training. The data points and error bars are obtained from ensembling 5 separate trains.}
    \label{fig:downsamp_loss}
\end{figure*}

To identify an optimal physics strength that can balance data fidelity with physics consistency, we analyzed the tradeoff between data loss and weighted physics loss ($\lambda \mathcal{L}_{physics}$) across varying magnitudes of $\lambda$. As shown in Figure \ref{fig:downsamp_loss}, in the low-weight regime ($\lambda \leq 10^{2}$), the weighted physics loss remains small, indicating that physical constraints have minimal influence on the optimization landscape, and the lack of physics regularization results in the physically inconsistent interpolations observed previously in Figure \ref{fig:downsamp_2}. As $\lambda$ increases toward the optimal value ($\sim 10^{4}$), the weighted physics loss starts to rise and intersect with the data loss. It indicates that the physical constraints have become active and start to shape the latent trajectory. In this regime, the data loss remains stable and flat, implying that the model successfully fulfills the governing PBM without violating the boundary data points. Therefore, in this case, a $\lambda$ around $10^{4}$ is optimal, with the two loss terms balanced at $10^{4}$. Beyond this critical threshold ($\lambda > 10^{6}$), the data loss shows a rapid increase as the optimization becomes dominated by the physics term at the expense of fitting the observed boundaries. The subsequent drop in weighted physics loss when $\lambda = 10^{10}$ suggests that the networks are biased toward a trivial solution that satisfies the ODEs by unphysically pushing the derivatives to zero.

Although the PIRNN successfully reconstructs the unobserved state trajectories using only the boundary points, the accuracy of the underlying kinetic parameters reveals the limitation of this extreme down-sampling, which is further quantified in Table \ref{tab:pirnn_param_downsamp}. Table \ref{tab:pirnn_param_downsamp} presents the PBM parameters estimated by the PIRNN across varying sampling densities, ranging from the boundary-only case to a sparse grid of 9 points per run. In the 2-point scenario, despite the accurate predicted trajectory shown in Figure \ref{fig:downsamp_2}, the estimated parameters deviate substantially from the reference values. This deviation indicates that while the physics constraints narrow the solution space to a physically feasible path, the problem remains ill-conditioned with respect to parameter identifiability, allowing the network to find alternative parameter combinations that satisfy the boundary states and the predicted trajectory. However, introducing a single intermediate data point significantly improves parameter recovery, retrieving all parameters except $k_g$ to their correct orders of magnitude. As the sampling density increases to 9 points per training run, the estimated parameters converge closely to the reference with $\sim$3.6\% mean squared error in logarithmic scale. This finding is promising for practical scenarios, where measuring $<10$ intermediate points is much more practical than sampling with high resolution.

\begin{table}[h]
\centering
\renewcommand{\arraystretch}{1.5}
\begin{adjustbox}{max width=\textwidth}
\begin{tabular}{lcccccc}
\toprule
\textbf{PBM Parameter} & $k_{b2}$ & $\alpha$ & $\beta$ & $k_{g}$ & $E_{ag}$ & $\gamma_{g}$ \\
\midrule
\textbf{Reference Value} & $6.000\times10^{3}$ & 2.080 & 0.713 & $2.730\times10^{5}$ & $4.130\times10^{4}$ & 1.240 \\
\midrule
\textbf{2 Points per Run} & $1.288(271)\times10^{2}$ & 3.901(46) & 1.954(76) & 0.307(19) & $2.608(183)\times10^{1} $ & 2.273(8) \\
\midrule
\textbf{3 Points per Run} & $3.344(53)\times10^{3}$ & 2.078(40) & 0.737(4) & $2.632(1687)\times10^{4}$ & $3.602(229)\times10^{4}$ & 1.211(36) \\
\midrule
\textbf{5 Points per Run} & $3.289(25)\times10^{3}$ & 1.983(21) & 0.726(2) & $5.397(2254)\times10^{4}$ & $3.861(137)\times10^{4}$ & 1.147(21) \\
\midrule
\textbf{9 Points per Run} & $5.250(19)\times10^{3}$ & 1.975(1) & 0.720(1) & $1.759(170)\times10^{5}$ & $4.054(29)\times10^{4}$ & 1.189(1) \\
\bottomrule
\end{tabular}
\end{adjustbox}
\vspace{3pt}
\caption{The PIRNN estimated PBM parameters under different down-sampling frequency. All parameters are optimized in logarithm scale to guarantee positivity. $\lambda$ is set to be $10^{4}$ as explained in the main context. The standard deviations are provided using parenthesis notation. The reference value are from previous work by \citet{Li2017ModelingCrystallization}. All models are successfully converged, and the convergence can be found in Figure \ref{fig:downsamp_model} in the SI. The re-evaluated ODE curves and the PIRNN predictions in the converged cases are also available as Figure \ref{fig:downsamp_3} to \ref{fig:downsamp_9} in the SI.}
\label{tab:pirnn_param_downsamp}
\end{table}

This investigation of sampling uncertainty demonstrates the dual utility of the PIRNN framework as both a dynamic interpolator and a parameter estimator. The PIRNN can bridge the unobserved gap between sparse measurements, allowing physically consistent state reconstruction even when measurements are limited. At the same time, although the state trends can be recovered correctly using only the start and end points, the kinetic parameters cannot be accurately retrieved, and it requires at least one additional intermediate observation to resolve the non-uniqueness inherent in the inverse problem. These results suggest that, while the PIRNN framework shows high robustness and effectiveness under the presence of sampling uncertainty, the temporal resolution is still crucial for reliable parameter estimation.

\section{Conclusions}
\label{sec:conc}
This paper systematically evaluated the robustness and effectiveness of PIRNNs in modeling batch crystallization under the effects of data non-idealities, including aleatoric and epistemic uncertainties. Using synthetic datasets with controlled levels of measurement noise, model mismatch, and data sparsity, we evaluated the generalization, physical consistency, and parameter interpretability of the PIRNN architecture. The results demonstrated that PIRNNs provide a robust and reliable approach to model crystallization dynamics under non-ideal conditions.

The inclusion of physics-based constraints is effective at alleviating the effects of aleatoric uncertainty. Although increasing data noise level established an irreducible error floor, the model’s predictive accuracy for underlying noise-free dynamics remained high and consistent. In addition, the framework consistently recovered physically meaningful parameters that closely match the reference values under noisy conditions, highlighting its ability to learn the underlying governing physical laws despite stochastic variations in training data. However, it must still be noted that the measurement data quality is crucial for learning accurate kinetics. In high-noise environments, PIRNNs act more like a robust state estimator by finding kinetically equivalent solutions that satisfy the dynamic constraints, even if the identification of a unique set of intrinsic parameters is limited by the data quality. 

In the presence of epistemic uncertainty introduced by a solubility model mismatch, incorporating an imperfect physics model as a soft constraint reveals a clear bias-variance tradeoff. Appropriately weighted physics constraints regularized the model training without ignoring the observed evidence required to correct the structural model error. Under this condition, the direct forward PIRNN prediction provides better predictive accuracy compared to the strict integration of the governing ODEs, highlighting the ability of the hybrid approach to bridge the gap between imperfect domain knowledge and experimental observations.

The analysis of sampling uncertainty demonstrates the framework's ability to reconstruct complex unobservable transient dynamics using a limited amount of information. Although boundary points alone are not sufficient to extract reliable physics, the inclusion of a small number of intermediate data points can significantly increase the accuracy of the learned kinetics.

This research validates the effectiveness of physics-informed machine learning as a powerful tool for developing robust predictive models of crystallization processes from imperfect data. The ability of the PIRNN framework to handle both random and systematic uncertainties offers a potential pathway toward more reliable process monitoring, optimization, and control. Future work will extend this framework to incorporate raw PAT data directly and explore adaptive weighting schemes to dynamically balance the data and physics fidelity.

\section*{Acknowledgments}
The authors thank the financial support from the Thomas A. Fanning Chair in Equity-Centered Engineering and the Georgia Tech College of Engineering ADVANCE Professorship. The authors are also grateful to Dr. Fernando Arrais R.D. Lima for the discussion of PBM and the method of moments implementation.

\bibliography{references}
\bibliographystyle{unsrtnat} 

\appendix
\newpage
\setcounter{section}{0}
\setcounter{page}{1}
\setcounter{figure}{0}
\setcounter{equation}{0}
\setcounter{table}{0}
\renewcommand{\thesection}{S\arabic{section}}
\renewcommand{\thepage}{s\arabic{page}}
\renewcommand{\thetable}{S\arabic{table}}
\renewcommand{\thefigure}{S\arabic{figure}}

\makeatletter
\let\@title\@empty
\makeatother

\begin{center}
    \textbf{\Large Supporting Information}\\\vspace{16pt} \textbf{\Large Modeling Batch Crystallization under Uncertainty Using Physics-informed Machine Learning} \\[1.5em]

    Dingqi Nai\textsuperscript{a}, Huayu Li\textsuperscript{b}, 
    Martha Grover\textsuperscript{a}, Andrew J. Medford\textsuperscript{a}$^*$ \\[1em]

    {
    \textsuperscript{a}School of Chemical \& Biomolecular Engineering\\
    Georgia Institute of Technology\\
    Atlanta, GA 30332\\[0.5em]

    \textsuperscript{b}Material and Analytical Sciences\\
    Boehringer Ingelheim Pharmaceuticals, Inc.\\
    Ridgefield, CT 06877\\
    }
\end{center}
\vspace{2em}

\section{Data Preparation}
\subsection{Synthetic Temperature Profiles}

\begin{figure*}[h!]
    \centering\includegraphics[keepaspectratio=true,scale=0.8]{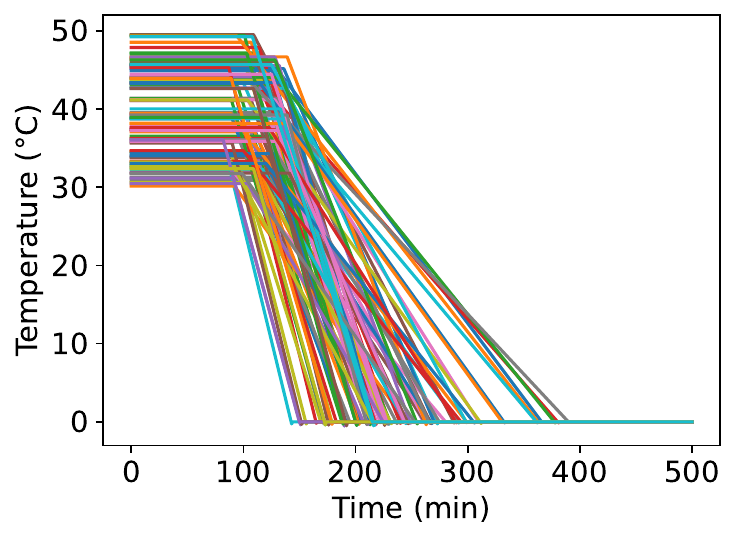}
    \caption{Temperature profile of the 100 synthetic batch cooling runs. Initial temperature, cooling rate, and plateau duration are randomly sampled from a uniform distribution.}
    \label{fig:temp_profile}
\end{figure*}

\begin{figure*}[h!]
    \centering\includegraphics[keepaspectratio=true,scale=0.8]{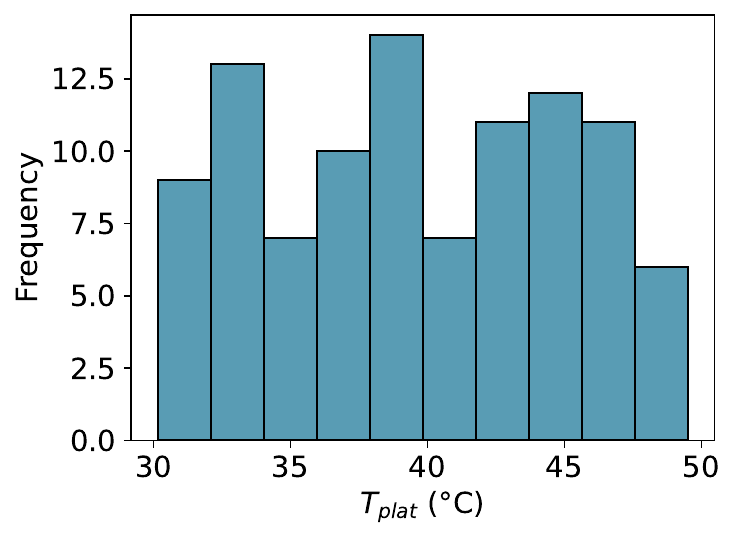}
    \caption{Distribution of the plateau temperature. 100 points are sampled from a uniform distribution from 30 to 50 $\unit{\celsius}$.}
    \label{fig:dist_tplat}
\end{figure*}

\begin{figure*}[h!]
    \centering\includegraphics[keepaspectratio=true,scale=0.8]{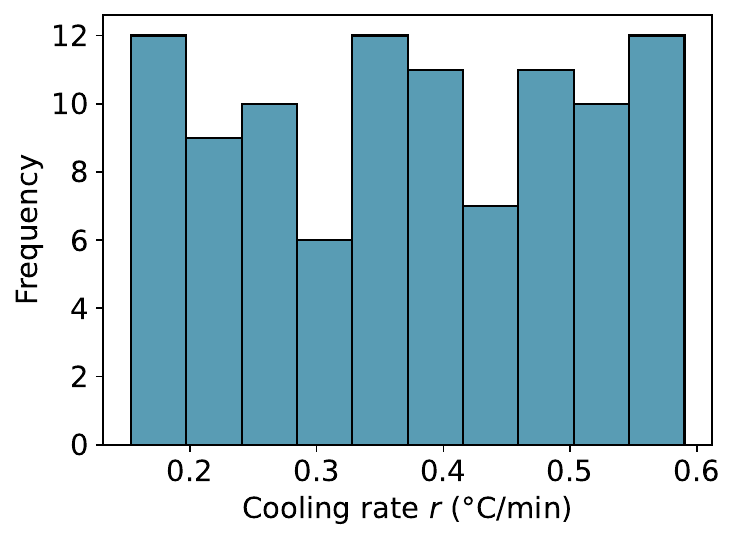}
    \caption{Distribution of the cooling rate. 100 points are sampled from a uniform distribution from 0.15 to 0.60 $\unit{\celsius}$/min.}
    \label{fig:dist_r}
\end{figure*}

\begin{figure*}[h!]
    \centering\includegraphics[keepaspectratio=true,scale=0.8]{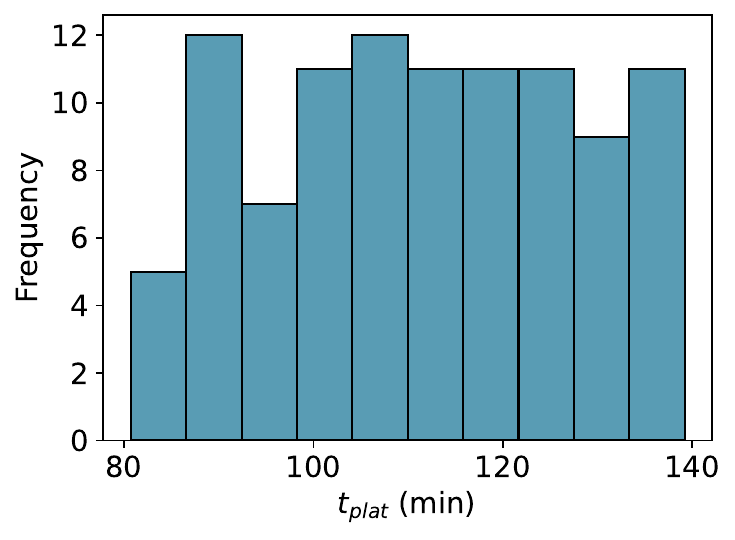}
    \caption{Distribution of the plateau duration. 100 points are sampled from a uniform distribution from 80 to 140 min.}
    \label{fig:dist_plat}
\end{figure*}

\clearpage
\newpage
\subsection{Synthetic Data} \label{subsec:SI_data}
The data is synthesized by forward solving PBM using method of moments 

\begin{figure*}[h!]
    \centering\includegraphics[keepaspectratio=true,scale=0.8]{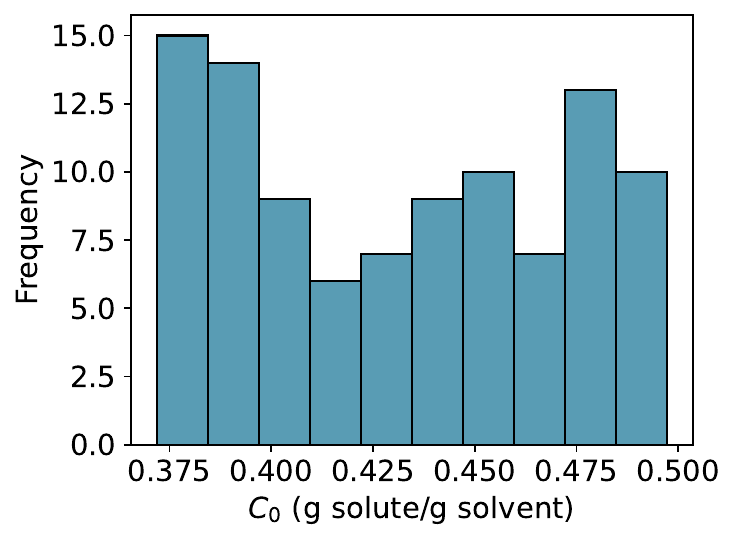}
    \caption{Distribution of the initial concentration used to generate the synthetic data. 100 points are sampled from a uniform distribution from 0.37 to 0.50 g solute/g solvent.}
    \label{fig:dist_c0}
\end{figure*}

\begin{figure*}[h!]
    \centering\includegraphics[keepaspectratio=true,scale=0.7]{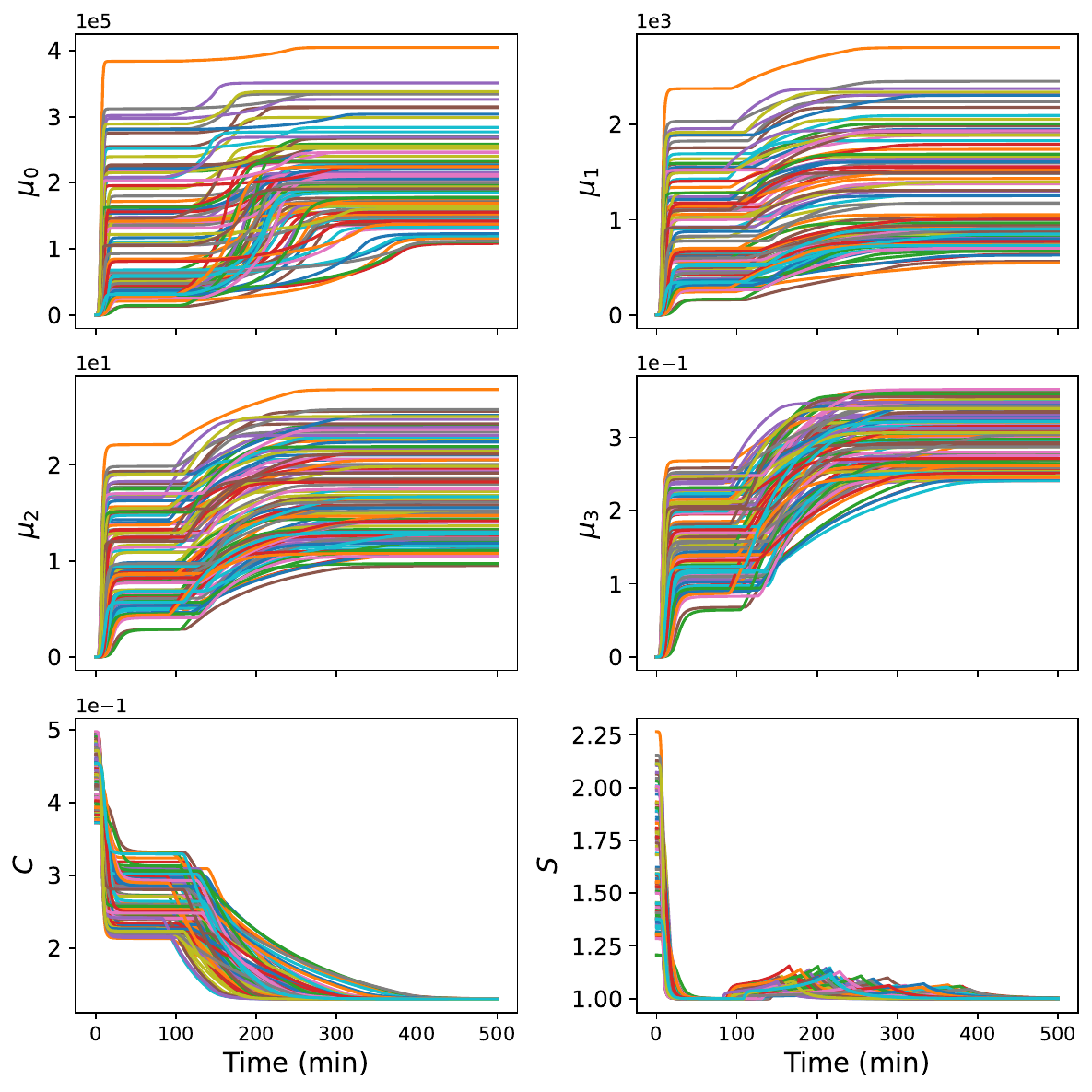}
    \caption{The synthetic noiseless data profiles. The supersaturation is also included for reference, but it is not included in the training.}
    \label{fig:data_profile_noiseless}
\end{figure*}

\begin{figure*}[h!]
    \centering\includegraphics[keepaspectratio=true,scale=0.7]{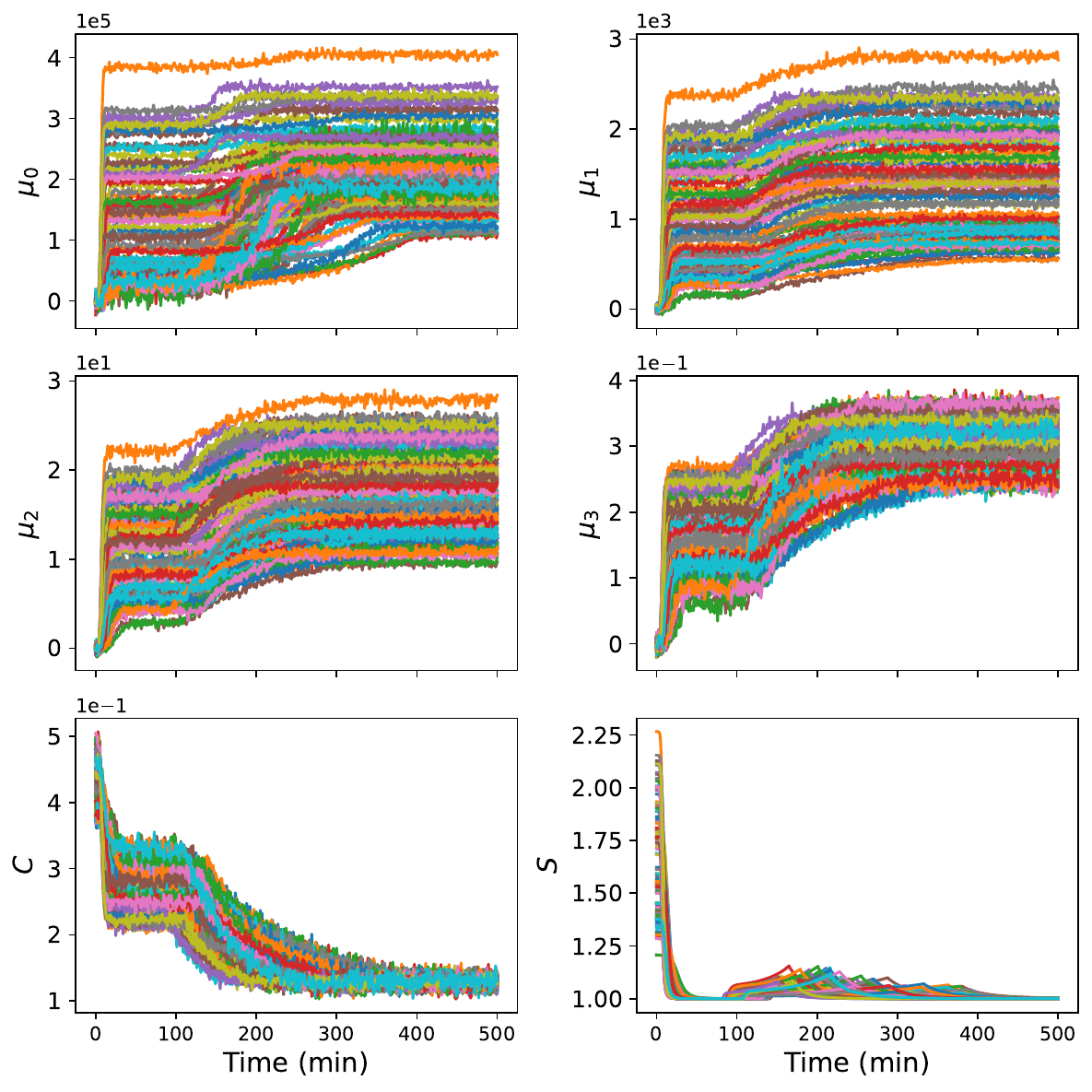}
    \caption{The synthetic 0.1 std noise level data profiles. The supersaturation is also included for reference, but it is not included in the training.}
    \label{fig:data_profile_01std}
\end{figure*}

\begin{figure*}[h!]
    \centering\includegraphics[keepaspectratio=true,scale=0.7]{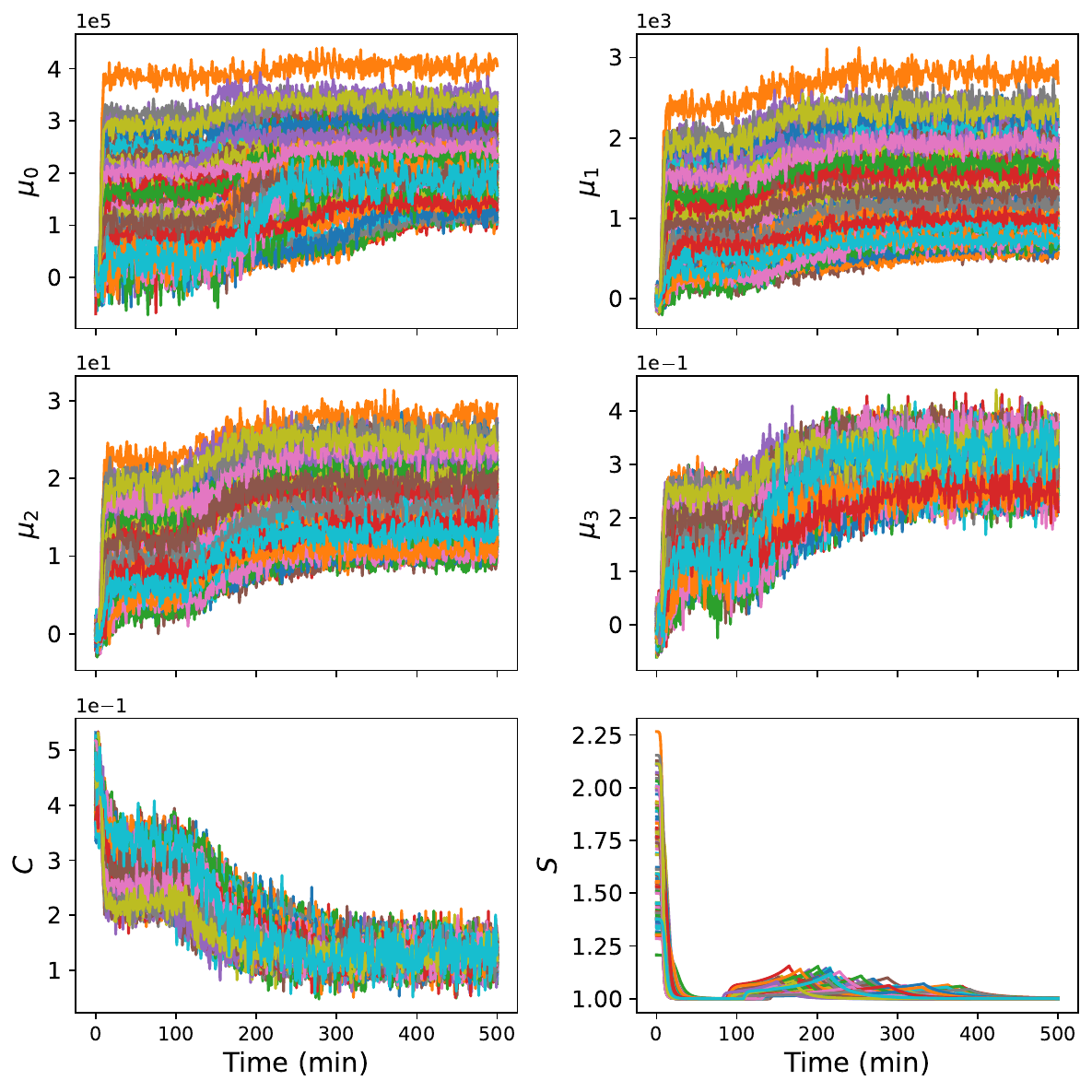}
    \caption{The synthetic 0.3 std noise level data profiles. The supersaturation is also included for reference, but it is not included in the training.}
    \label{fig:data_profile_03std}
\end{figure*}

\begin{figure*}[h!]
    \centering\includegraphics[keepaspectratio=true,scale=0.7]{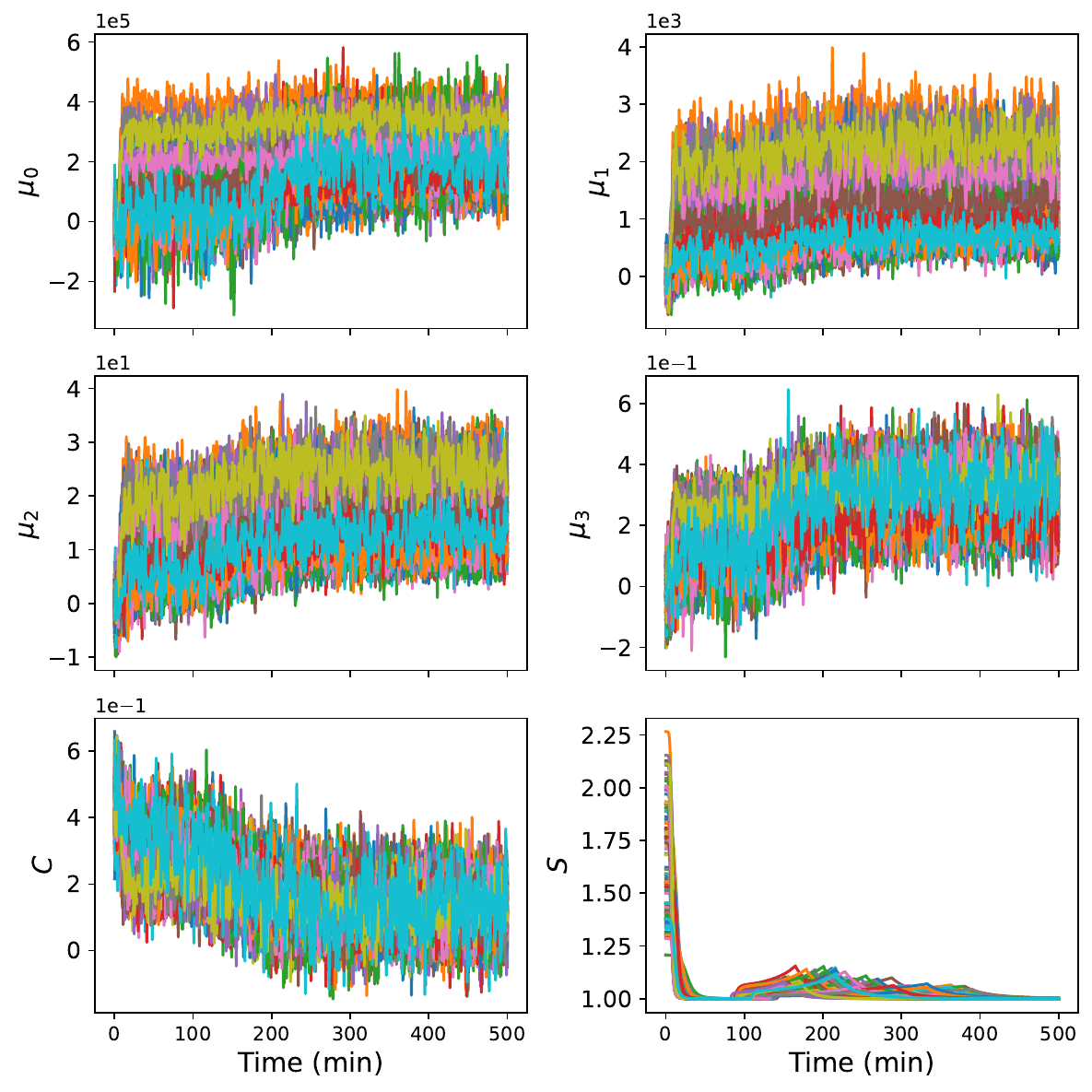}
    \caption{The synthetic 1.0 std noise level data profiles. The supersaturation is also included for reference, but it is not included in the training.}
    \label{fig:data_profile_10std}
\end{figure*}

\clearpage
\newpage
\subsection{NN Architecture}
The specific designs and hyperparameters used for the network architecture and training configuration are summarized in Table \ref{tab:pirnn_hyperparameters}. The input and output dimensions are selected to match the number of state and control variables. 64 hidden units with 2 LSTM layers were selected to ensure sufficient representation capability while avoiding too much complexity; 64 hidden units are also chosen for hardware efficiency to match binary logic, and it is also a commonly used empirical start point. The 0.2 dropout rate is selected within the commonly used range, as shown in the Pytorch documentation example. The lower dropout is used also because RNN/LSTM is known to be more sensitive to dropout than CNN. Softplus activation is selected because of the smoothness and the guaranty of nonnegative values. The initial noise scale and transition threshold are selected as 0.1 since all state variables are normalized in the range 0 to 1. In this case, 0.1 (10\%) can be considered as a fairly large discrepancy. Adam is a widely used optimizer, and 1$\times 10^{-3}$ is the default learning rate for the Pytorch Adam optimizer. 

These settings were kept fixed in all case studies. We did not conduct a careful hyperparameter search as the goal of this paper is to demonstrate the general robustness and validity of the PIRNN framework under controlled uncertainties. Systematic hyperparameter optimization may improve the results, but should not affect the conclusions.
\begin{table}[h]
    \centering
    \renewcommand{\arraystretch}{1.2}
    \begin{adjustbox}{max width=\textwidth}
    \begin{tabular}{lcl}
        \toprule
        \textbf{Hyperparameter} & \textbf{Value} & \textbf{Description} \\ 
        \midrule
        Input Dimension     & 6   & State variables ($\mathbf{x}$) + Control input ($T$) \\
        Output Dimension    & 5   & Predicted state variables \\
        Hidden State Size   & 64  & Number of features in the LSTM hidden state \\
        Number of Layers    & 2   & Number of stacked LSTM layers \\
        Dropout Rate        & 0.2 & Dropout probability applied after the first LSTM layer \\
        Activation Function & Softplus & Applied to final output to ensure non-negativity \\
        Normalization       & Batch Norm 1D & Applied to the LSTM output before the decoder \\
        Initial Noise Scale $\eta$ & 0.1 & Initial value for the learnable noise parameter \\
        Transition Threshold $\delta$ & 0.1 & Transition threshold of Huber loss \\
        Optimizer           & Adam & Optimized for both model weights and PBM parameters \\
        Base Learning Rate  & 1$\times 10^{-3}$ & Initial learning rate for the Adam optimizer \\
        LR Scheduler        & Cosine Annealing & Decay learning rate to $1 \times 10^{-7}$ over total epochs \\
        \bottomrule
    \end{tabular}
    \end{adjustbox}
    \vspace{3pt}
    \caption{The network hyperparameters and training configurations.}
    \label{tab:pirnn_hyperparameters}
\end{table}

\clearpage
\newpage
\section{Signal Noise}

As training data size increases, the PIRNN prediction aligns closer to the noiseless reference, indicating a better generalization of the network itself. The reason the model generally performs better in $\mu_2$, $\mu_3$ and $C$ than in $\mu_0$ and $\mu_1$ is that the PIRNN does not treat each state variable ``equally'' during optimization. Although all state variables are normalized to a range from 0 to 1, their derivatives still have different orders of magnitude. Therefore, due to their different ease of fitting and scale of time derivatives, a limited generalized model may prioritize the fitting of highly correlated and relatively easy to fit state variables, which are $\mu_2$, $\mu_3$, and $C$ in this case. In all cases except the high-noise case, the re-evaluated ODE curves are consistent and accurate, suggesting a high reliability of the PIRNN estimated kinetic parameters. In the high-noise case, a discrepancy is observed between the re-evaluated $\mu_0$ and the ground-truth reference, indicating a degradation in the extracted kinetic quality. The discrepancy decreases from $\mu_0$ to $\mu_1$ to $\mu_2$, suggesting a possible error accumulation in the ODE integration. In the high-noise case, although the prediction of the re-evaluated ODE shows obvious discrepancy, the PIRNN prediction remains accurate, highlighting the advantage of possible dual usage of hybrid modeling.

\begin{figure*}[h!]
    \centering\includegraphics[keepaspectratio=true,scale=0.54]{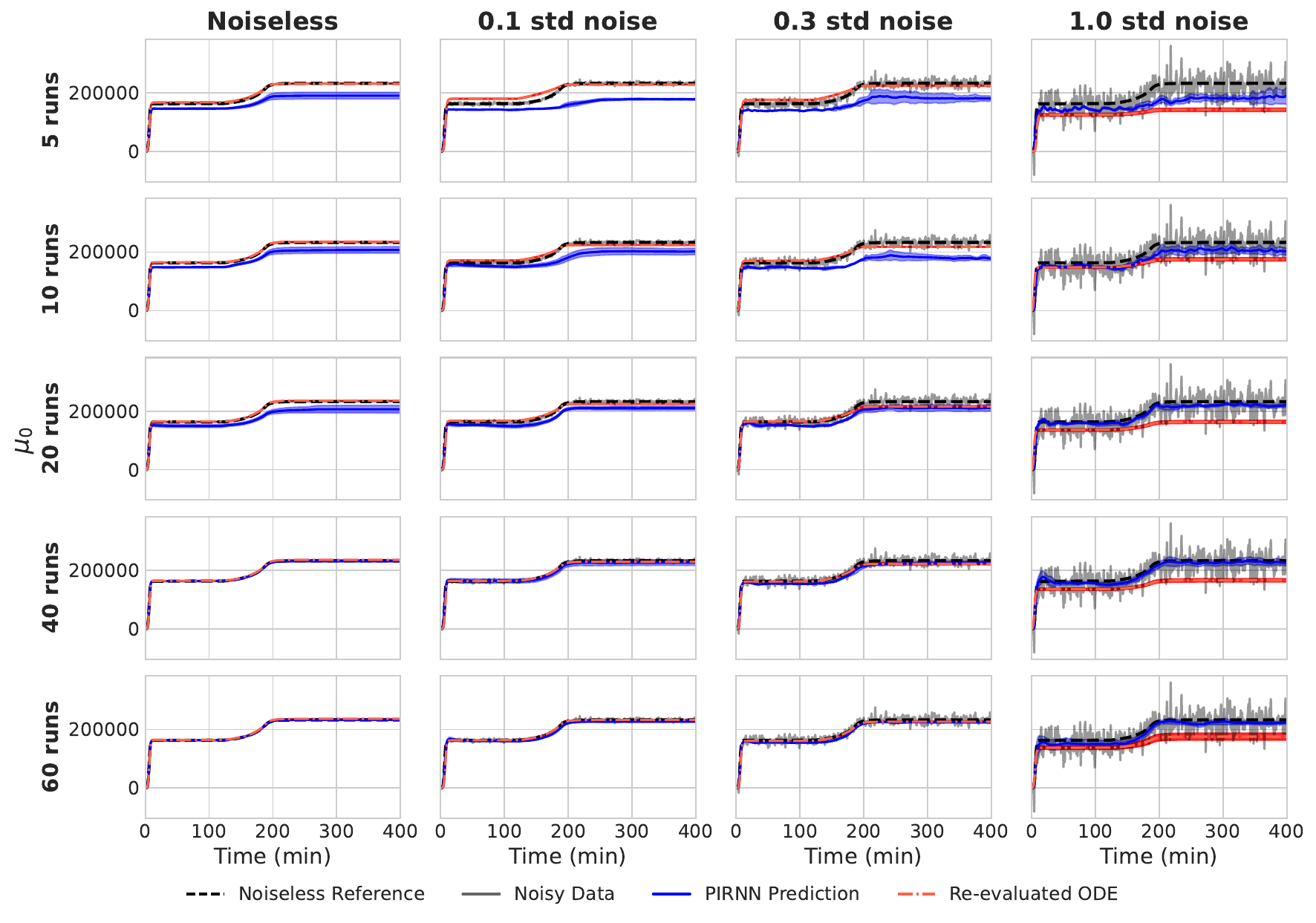}
    \caption{$\mu_0$ prediction of a set of test data using 5, 10, 20, 40, and 60 training runs. The 95\% confidence intervals are provided. The original noisy data and ground-truth noiseless reference are shown for each case.}
    \label{fig:test_pred_mu0}
\end{figure*}

\begin{figure*}[h!]
    \centering\includegraphics[keepaspectratio=true,scale=0.54]{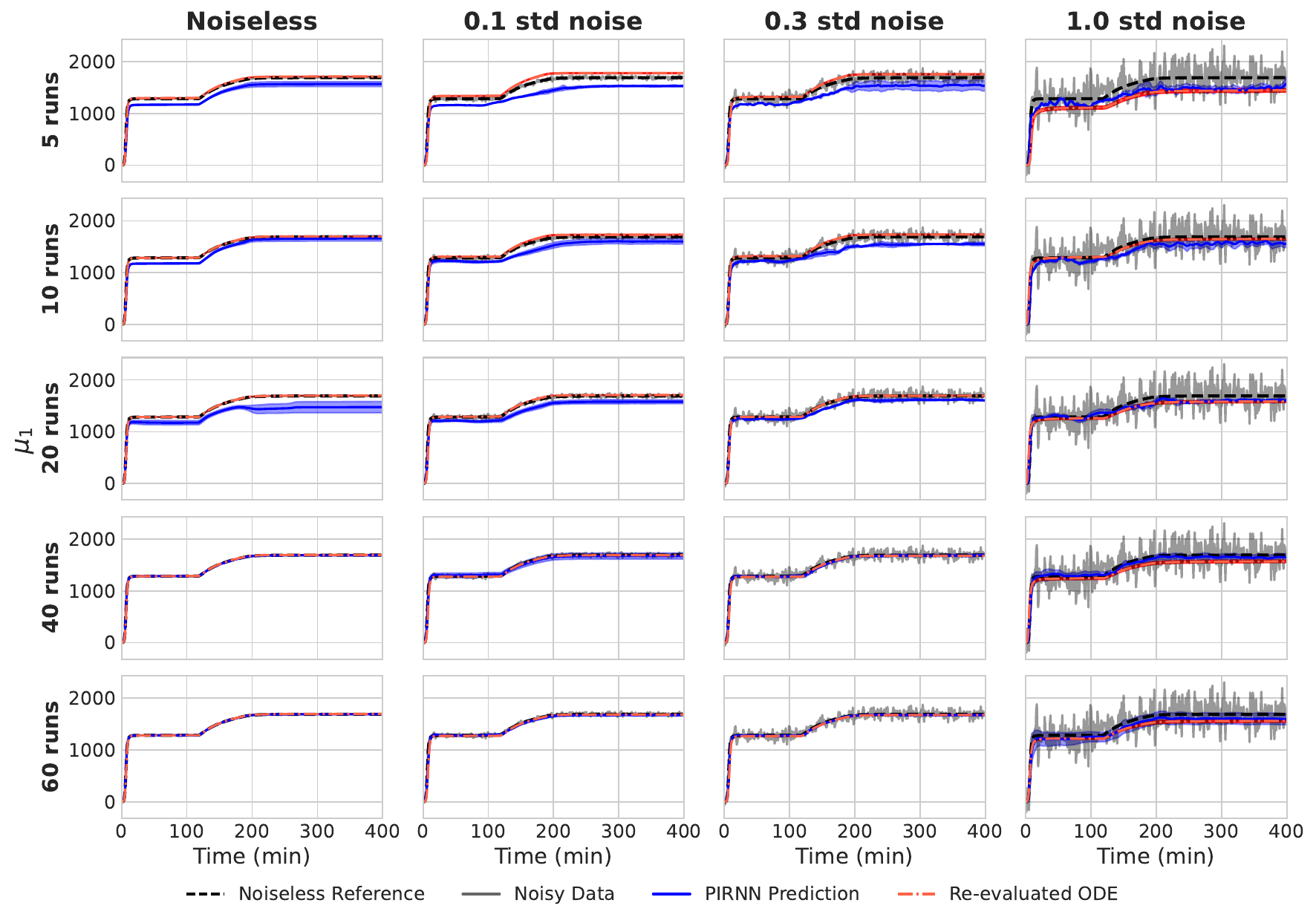}
    \caption{$\mu_1$ prediction of a set of test data using 5, 10, 20, 40, and 60 training runs. The 95\% confidence intervals are provided. The original noisy data and ground-truth noiseless reference are shown for each case.}
    \label{fig:test_pred_mu1}
\end{figure*}

\begin{figure*}[h!]
    \centering\includegraphics[keepaspectratio=true,scale=0.54]{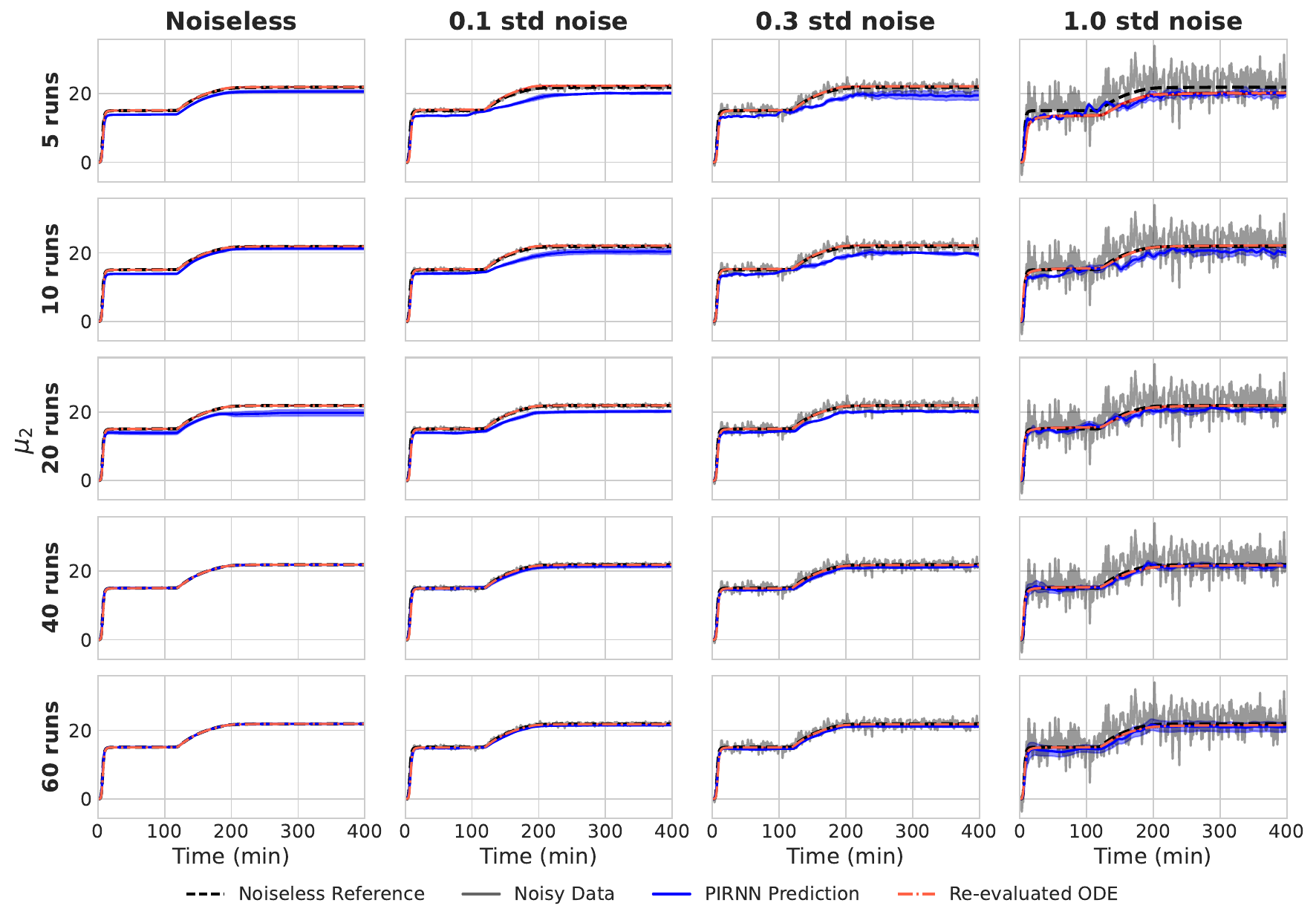}
    \caption{$\mu_2$ prediction of a set of test data using 5, 10, 20, 40, and 60 training runs. The 95\% confidence intervals are provided. The original noisy data and ground-truth noiseless reference are shown for each case.}
    \label{fig:test_pred_mu2}
\end{figure*}

\begin{figure*}[h!]
    \centering\includegraphics[keepaspectratio=true,scale=0.54]{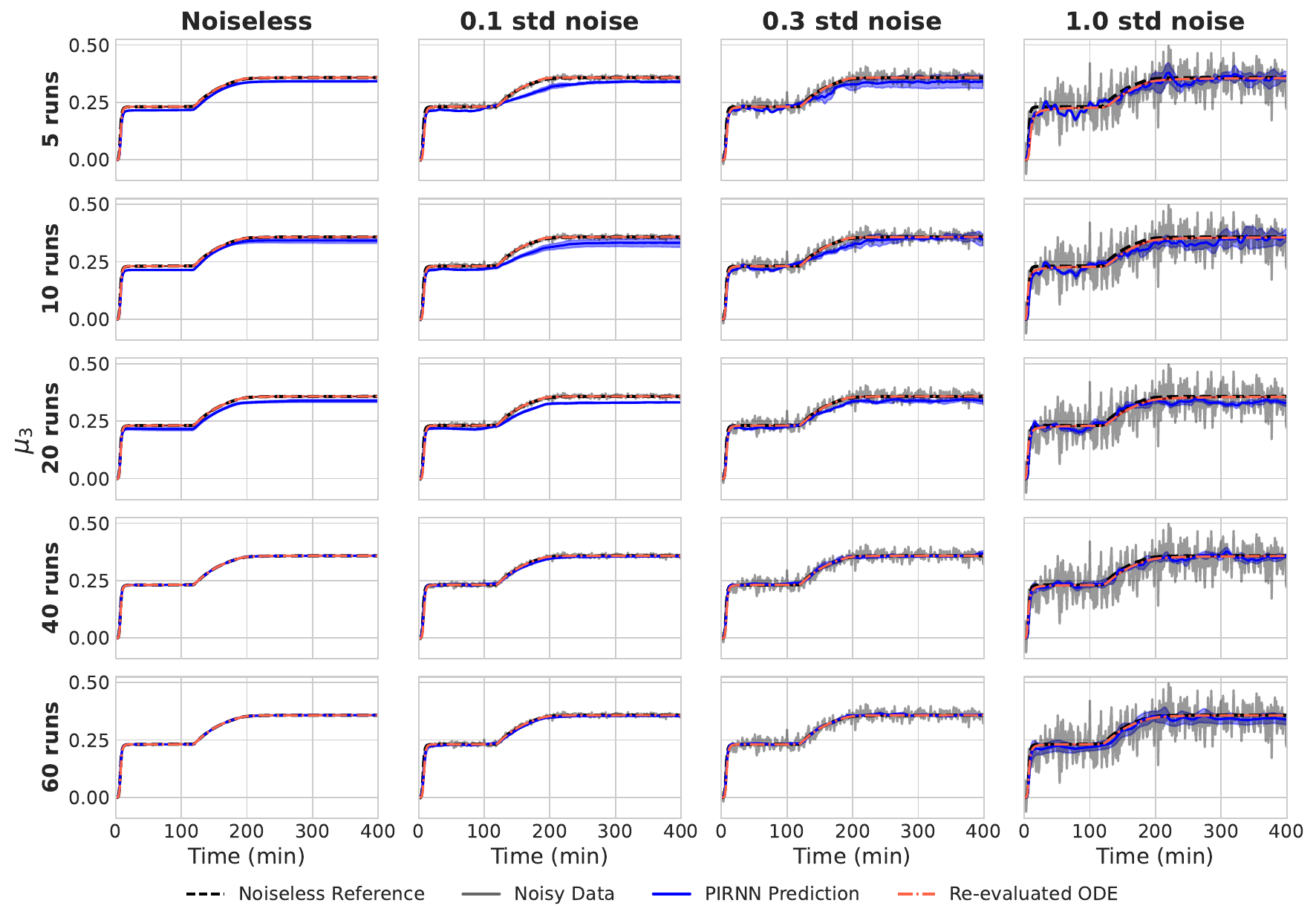}
    \caption{$\mu_3$ prediction of a set of test data using 5, 10, 20, 40, and 60 training runs. The 95\% confidence intervals are provided. The original noisy data and ground-truth noiseless reference are shown for each case.}
    \label{fig:test_pred_mu3}
\end{figure*}

\clearpage
\newpage

\section{Solubility Shift}

\begin{figure*}[h!]
    \centering\includegraphics[keepaspectratio=true,scale=0.54]{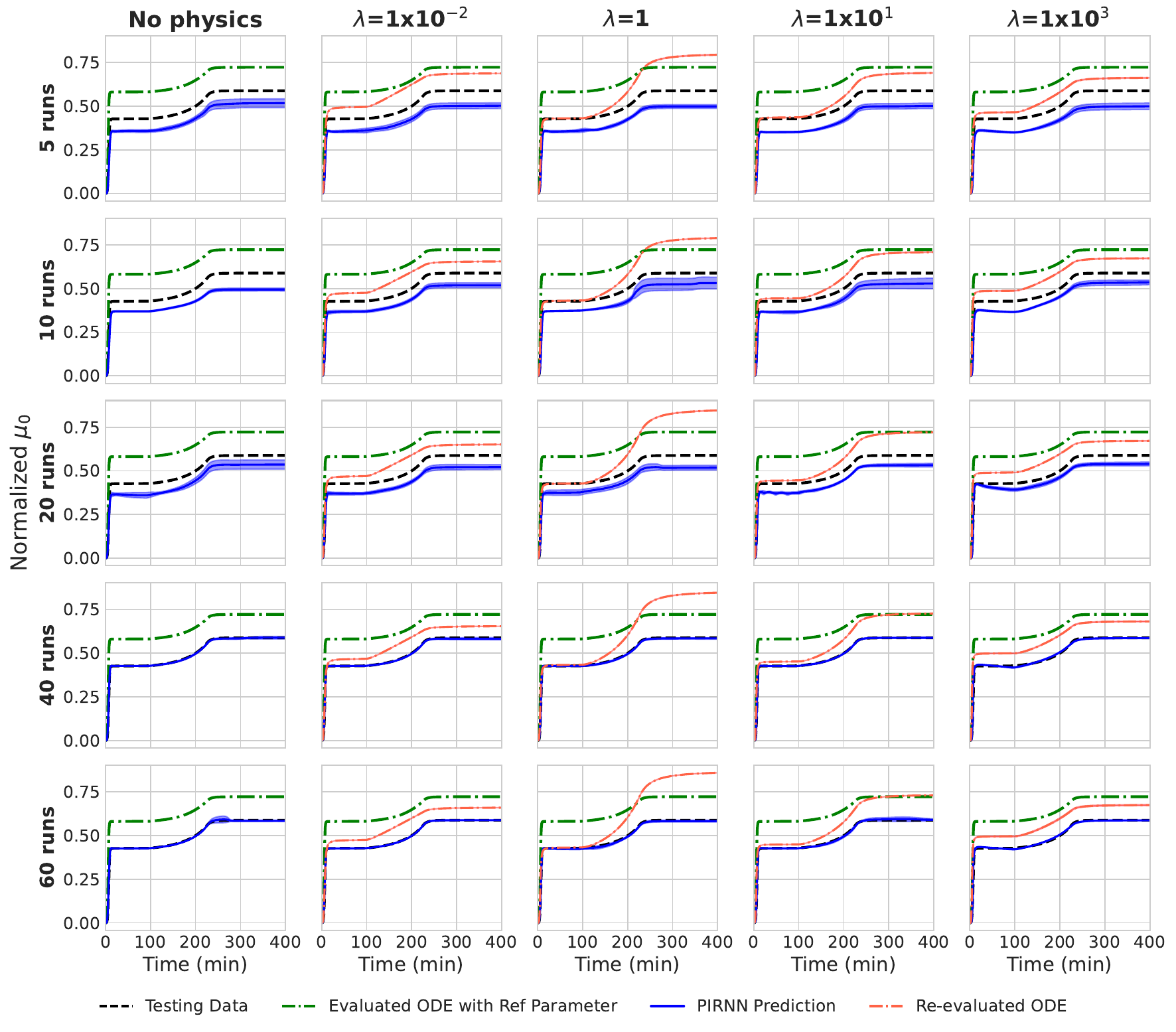}
    \caption{Normalized $\mu_0$ prediction of a set of test data using 5, 10, 20, 40, and 60 training runs at different $\lambda$ when 10\% constant solubility shift is applied. Re-evaluated ODE curves are not shown for no physics case since the PBM parameters are not included in optimization when physics loss is not included. The ODE curves are evaluated using two different sets of parameter: the reference values (green) and the PIRNN optimized values (red). The 95\% confidence intervals are shown.}
    \label{fig:test_solushift_mu0}
\end{figure*}

\begin{figure*}[h!]
    \centering\includegraphics[keepaspectratio=true,scale=0.54]{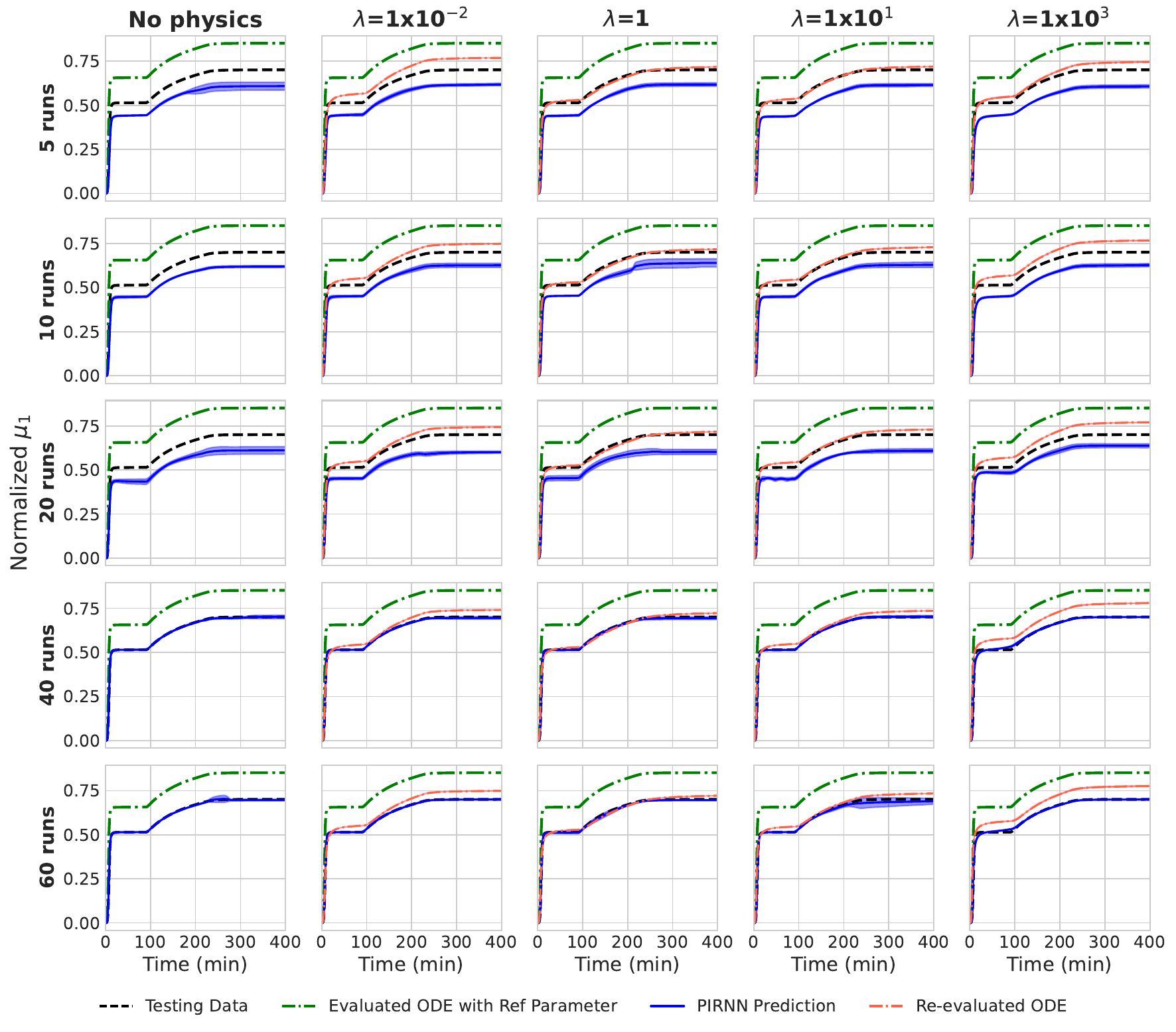}
    \caption{Normalized $\mu_1$ prediction of a set of test data using 5, 10, 20, 40, and 60 training runs at different $\lambda$ when 10\% constant solubility shift is applied. Re-evaluated ODE curves are not shown for no physics case since the PBM parameters are not included in optimization when physics loss is not included. The ODE curves are evaluated using two different sets of parameter: the reference values (green) and the PIRNN optimized values (red). The 95\% confidence intervals are shown.}
    \label{fig:test_solushift_mu1}
\end{figure*}

\begin{figure*}[h!]
    \centering\includegraphics[keepaspectratio=true,scale=0.54]{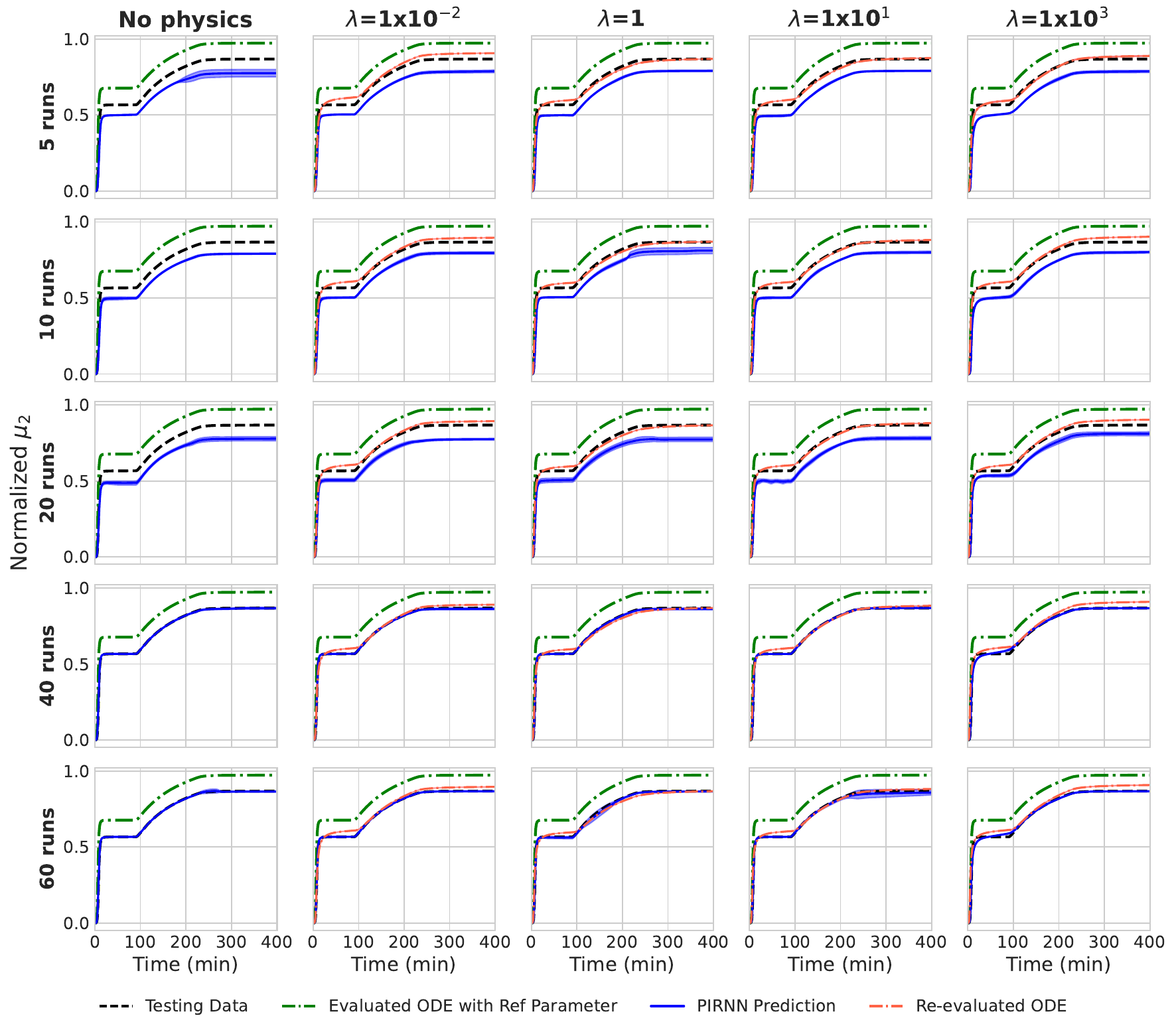}
    \caption{Normalized $\mu_2$ prediction of a set of test data using 5, 10, 20, 40, and 60 training runs at different $\lambda$ when 10\% constant solubility shift is applied. Re-evaluated ODE curves are not shown for no physics case since the PBM parameters are not included in optimization when physics loss is not included. The ODE curves are evaluated using two different sets of parameter: the reference values (green) and the PIRNN optimized values (red). The 95\% confidence intervals are shown.}
    \label{fig:test_solushift_mu2}
\end{figure*}

\begin{figure*}[h!]
    \centering\includegraphics[keepaspectratio=true,scale=0.54]{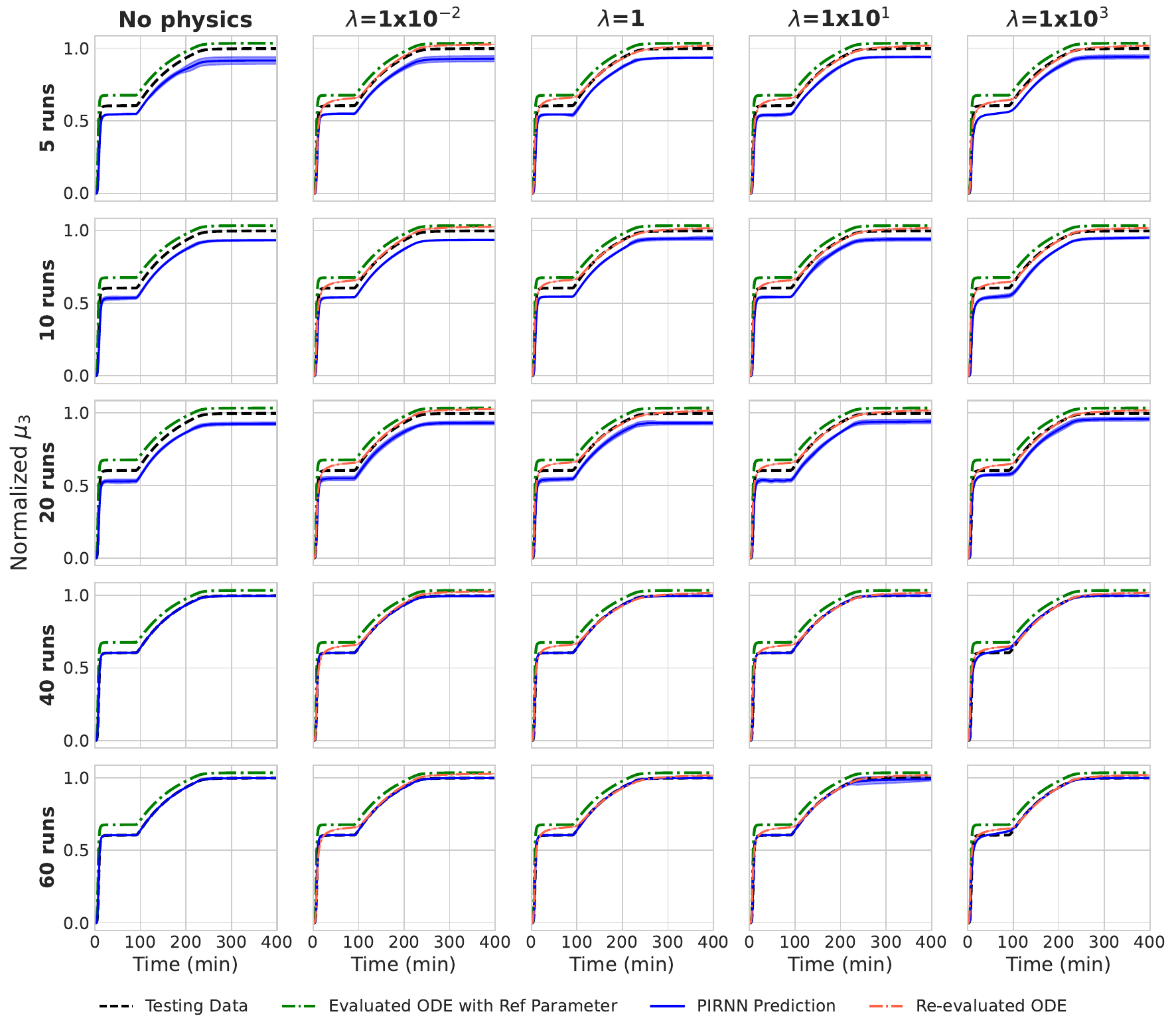}
    \caption{Normalized $\mu_3$ prediction of a set of test data using 5, 10, 20, 40, and 60 training runs at different $\lambda$ when 10\% constant solubility shift is applied. Re-evaluated ODE curves are not shown for no physics case since the PBM parameters are not included in optimization when physics loss is not included. The ODE curves are evaluated using two different sets of parameter: the reference values (green) and the PIRNN optimized values (red). The 95\% confidence intervals are shown.}
    \label{fig:test_solushift_mu3}
\end{figure*}

\clearpage
\newpage
\section{Sampling Frequency Limitations} \label{subsec:SI_downsamp}

High parity between the finite difference estimated time derivatives and the learned PBM parameters calculated time derivatives suggests that the model loss is successfully converged in this case. Therefore, the discrepancy between the learned parameter values and the reference is likely a result of kinetically equivalent parameter sets of the high inter-correlation between PBM parameters.

\begin{figure*}[h!]
    \centering\includegraphics[keepaspectratio=true,scale=0.7]{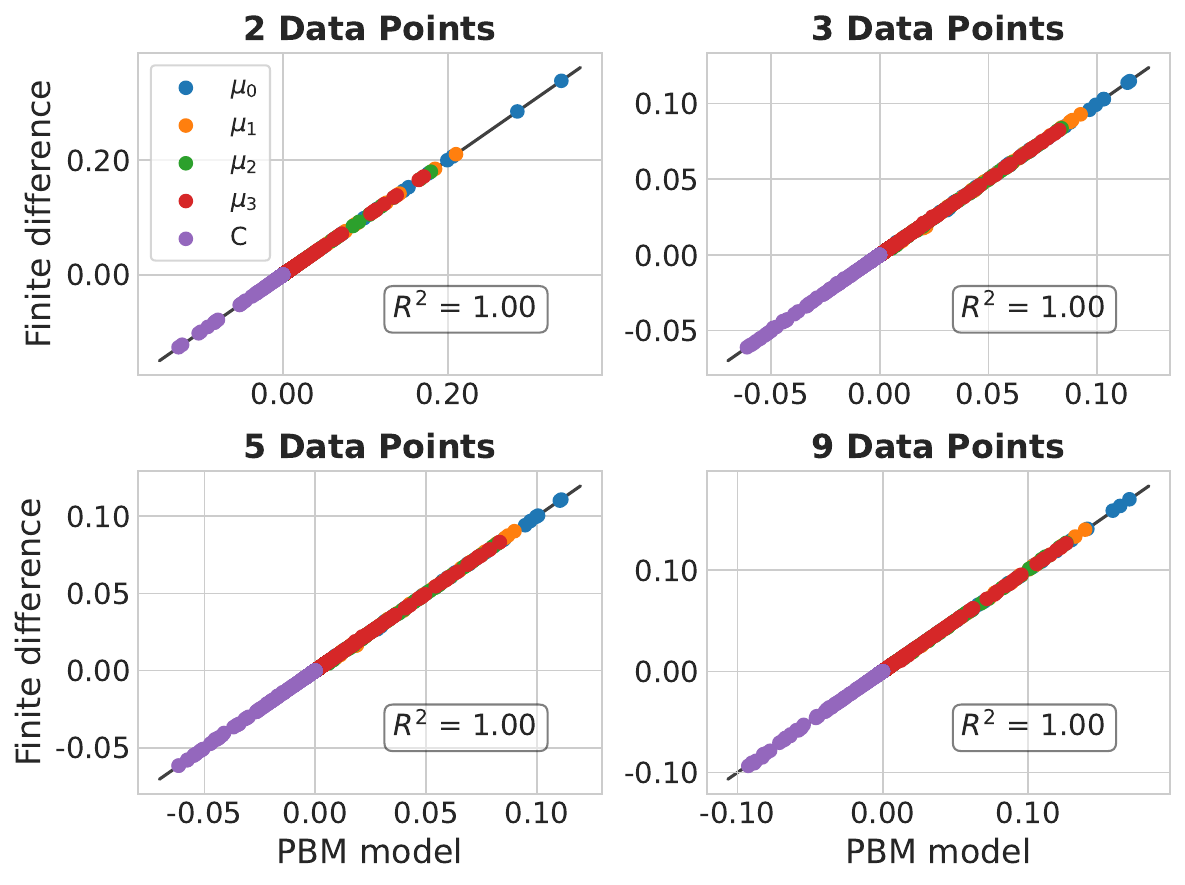}
    \caption{Model convergence between the finite difference estimated time derivatives and the time derivatives calculated with the PIRNN retrieved PBM model parameters in Table \ref{tab:pirnn_param_downsamp}. }
    \label{fig:downsamp_model}
\end{figure*}

\begin{figure*}[h!]
    \centering\includegraphics[keepaspectratio=true,scale=0.6]{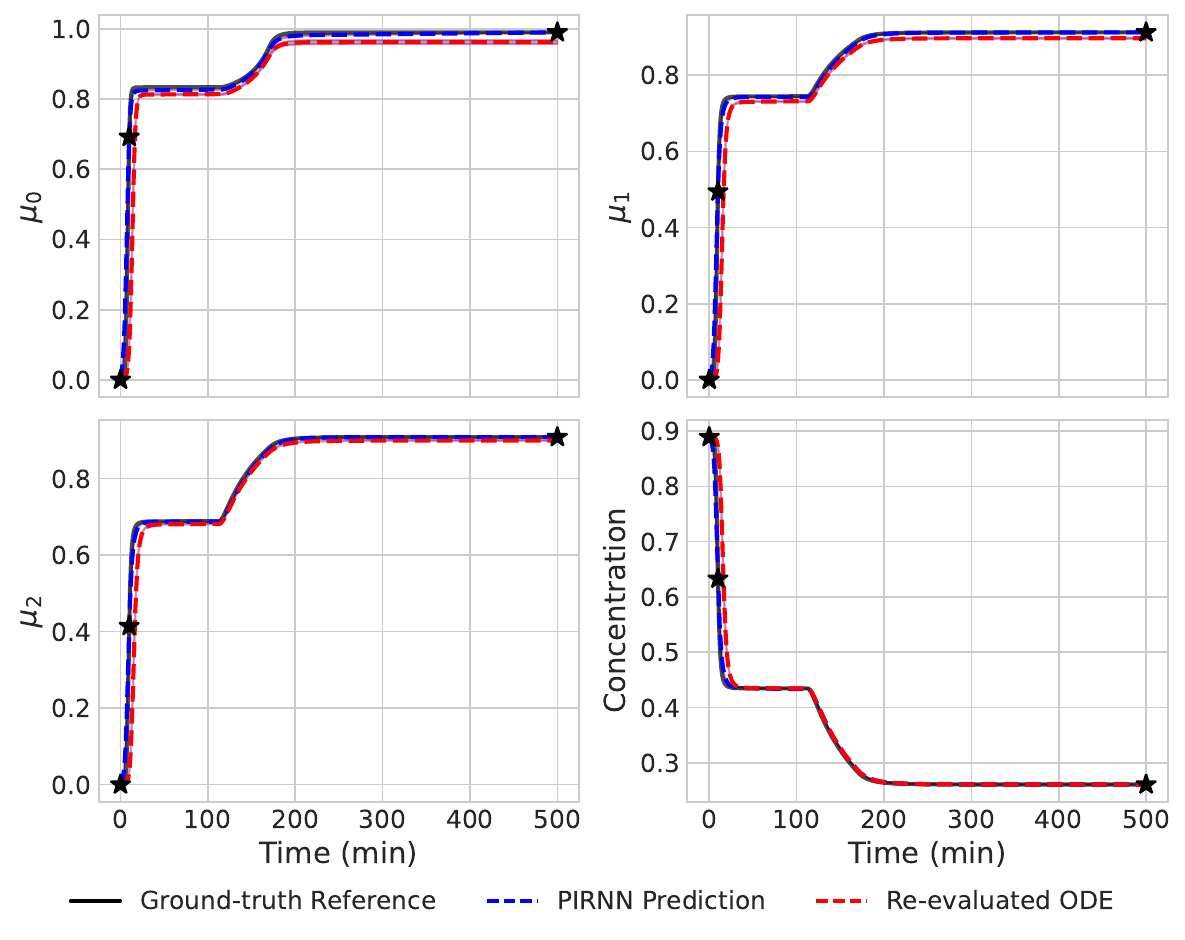}
    \caption{The PIRNN predicted trajectories and the evaluated ODE trajectories using PIRNNs estimated PBM parameters for normalized $\mu_0$, $\mu_1$, $\mu_2$, and $C$ when using 3 data points per run. The ground-truth references are shown as black solid lines.}
    \label{fig:downsamp_3}
\end{figure*}

\begin{figure*}[h!]
    \centering\includegraphics[keepaspectratio=true,scale=0.6]{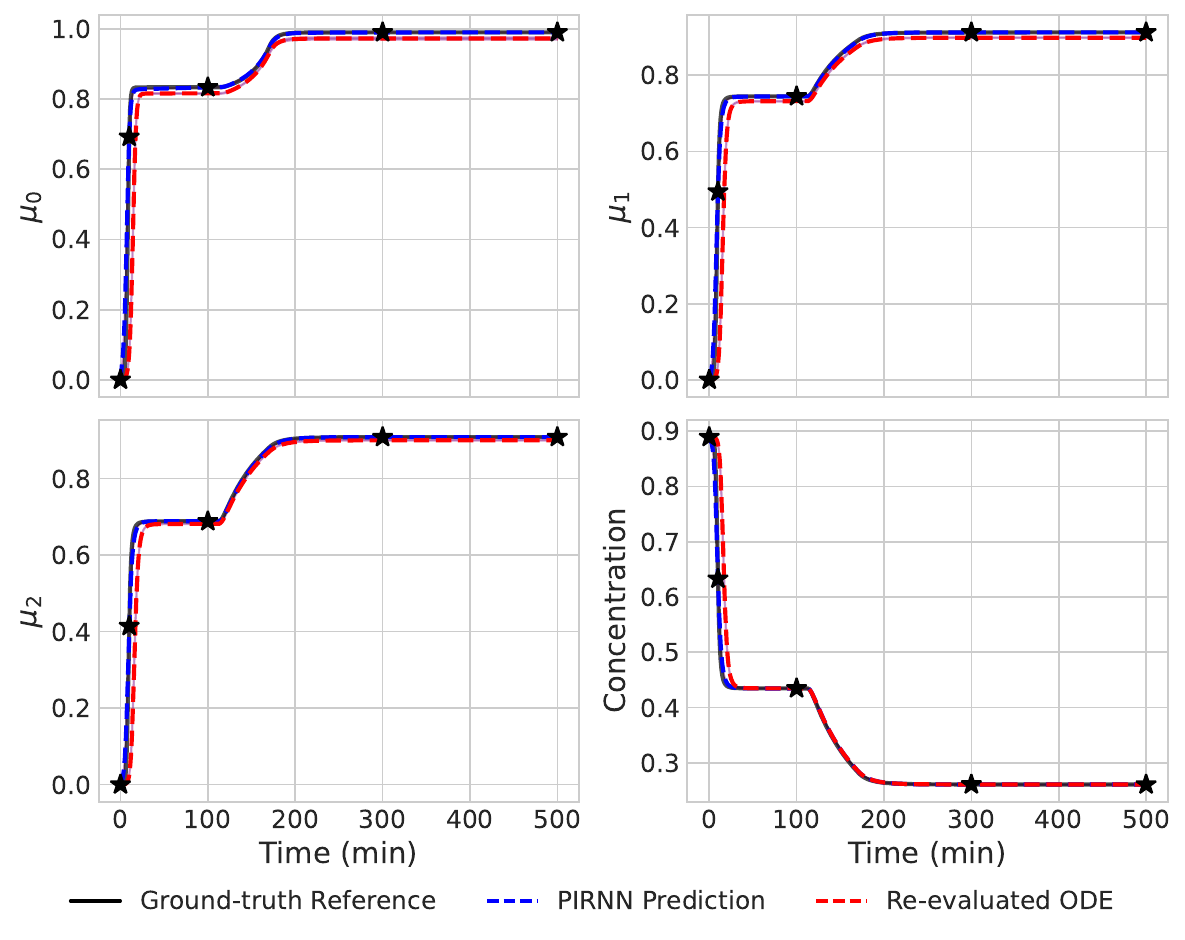}
    \caption{The PIRNN predicted trajectories and the evaluated ODE trajectories using PIRNNs estimated PBM parameters for normalized $\mu_0$, $\mu_1$, $\mu_2$, and $C$ when using 5 data points per run. The ground-truth references are shown as black solid lines.}
    \label{fig:downsamp_5}
\end{figure*}

\begin{figure*}[h!]
    \centering\includegraphics[keepaspectratio=true,scale=0.6]{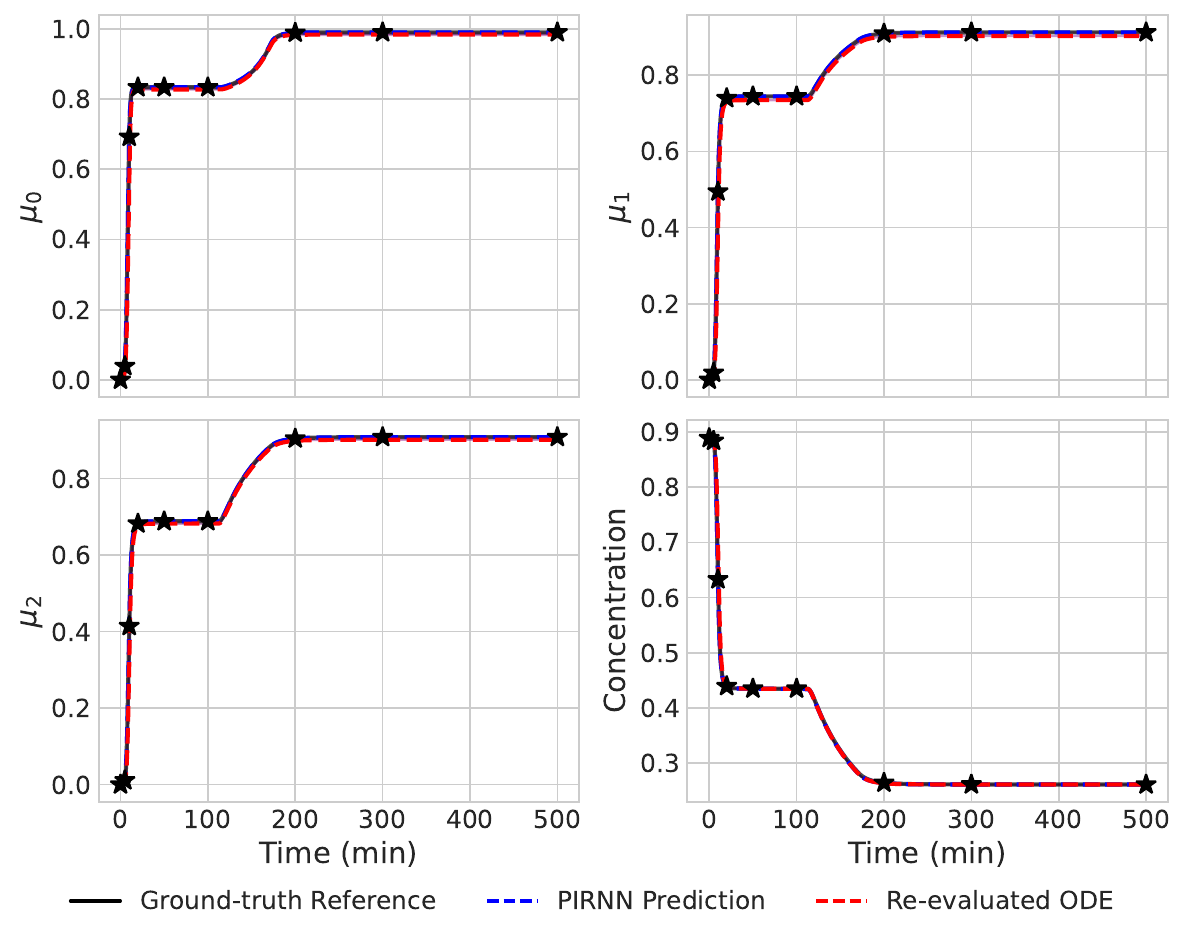}
    \caption{The PIRNN predicted trajectories and the evaluated ODE trajectories using PIRNNs estimated PBM parameters for normalized $\mu_0$, $\mu_1$, $\mu_2$, and $C$ when using 9 data points per run. The ground-truth references are shown as black solid lines.}
    \label{fig:downsamp_9}
\end{figure*}

\clearpage
\newpage
\section{Code Availability}

\end{document}